\documentclass[]{aa}  

\usepackage{txfonts,epsfig,graphicx,natbib,url,twoopt}
\usepackage{paralist}
\usepackage{booktabs,xspace}
\usepackage{amsmath}
\usepackage{tikz} 
\usepackage{amsmath}
\usepackage{amssymb}
\usepackage{graphicx}

\bibpunct{(}{)}{;}{a}{}{,}

\usepackage[hidelinks=true,breaklinks=true]{hyperref} 
\usepackage{natbib,twoopt}
\bibpunct{(}{)}{;}{a}{}{,} 
\makeatletter
 \newcommandtwoopt{\citeads}[3][][]{%
   \href{http://adsabs.harvard.edu/abs/#3}%
        {\def\hyper@linkstart##1##2{}%
         \let\hyper@linkend\@empty\citealp[#1][#2]{#3}}
	}
 \newcommandtwoopt{\citepads}[3][][]{%
   \href{http://adsabs.harvard.edu/abs/#3}%
        {\def\hyper@linkstart##1##2{}%
         \let\hyper@linkend\@empty\citep[#1][#2]{#3}}
	}
 \newcommandtwoopt{\citetads}[3][][]{%
   \href{http://adsabs.harvard.edu/abs/#3}%
        {\def\hyper@linkstart##1##2{}%
         \let\hyper@linkend\@empty\citet[#1][#2]{#3}}
	}
 \newcommandtwoopt{\citeyearads}[3][][]{%
   \href{http://adsabs.harvard.edu/abs/#3}%
        {\def\hyper@linkstart##1##2{}%
         \let\hyper@linkend\@empty\citeyear[#1][#2]{#3}}
	}
\makeatother

\newcommand{\sat}[1]{{\em #1}}
\newcommand{\G}[1]{\sat{Gaia\,#1}\xspace}
\newcommand{\Gaia}{\sat{Gaia}\xspace}
\newcommand{\NJ}{\sat{Nano-JASMINE}\xspace}
\newcommand{\TychoTwo}{\sat{Tycho-2}\xspace}
\newcommand{\Tycho}{\sat{Tycho}\xspace}
\newcommand{\Hipparcos}{\sat{Hipparcos}\xspace}
\newcommand{\Hip}{\sat{Hip}\xspace}
\newcommand{\AGISLab}{\textsc{agislab}\xspace}
\newcommand{\AGIS}{\textsc{agis}\xspace}
\newcommand{\SMOK}{\textsc{smok}\xspace}
\newcommand{\aposteriori}{{\em a~posteriori}\xspace}
\newcommand{\apriori}{{\em a~priori}\xspace}
\newcommand{\HTPM}{{\textsc{htpm}}\xspace}
\newcommand{\XHIP}{{\textsc{xhip}}\xspace}

\renewcommand{\vec}{\boldsymbol} 

\begin{document} 
	\title{Joint astrometric solution of \Hipparcos and \Gaia}
	\subtitle{A recipe for the Hundred Thousand Proper Motions project}
	\author{
		Daniel Michalik\inst{1}
			\and
		Lennart Lindegren\inst{1}
			\and
		David Hobbs\inst{1}
			\and
		Uwe Lammers\inst{2}
	}

	\institute{
		Lund Observatory, Lund University, Box 43, SE-22100 Lund, Sweden\\
		\email{daniel.michalik@astro.lu.se, lennart@astro.lu.se, david@astro.lu.se} 
		\and
		European Space Agency (ESA/ESAC), P.O. Box 78, ES-28691 Villanueva de la Ca\~{n}ada, Madrid, Spain\\
		\email {uwe.lammers@sciops.esa.int}
	}

	\date{\today}

	\abstract
	        {The first release of astrometric data from \Gaia is expected in 2016. It will contain 
	        the mean stellar positions and magnitudes from the first year of observations. 
	        For more than 100\,000 stars in common with the \Hipparcos
	        Catalogue it will be possible to compute very accurate proper motions due to the time
	        difference of about 24 years between the two missions. This Hundred Thousand Proper Motions
	        (\HTPM) project is planned to be part of the first release. }
		{Our aim is to investigate how early \Gaia data can be optimally combined with information from 
		the \Hipparcos Catalogue in order to provide the most accurate and reliable results for \HTPM.}
		{The Astrometric Global Iterative Solution (\AGIS) was developed to compute the astrometric core solution
		based on the \Gaia observations and will be used for all releases of astrometric data from \Gaia. We adapt 
		\AGIS to process \Hipparcos data in addition to \Gaia observations, and use simulations to verify and study the 
		joint solution method.}
		{For the \HTPM stars we predict proper motion accuracies between 14 and 134~$\mu$as~yr$^{-1}$, depending on stellar magnitude and
		amount of \Gaia data available. Perspective effects will be 
		important for a significant number of \HTPM stars, and in order to treat these effects accurately 
		we introduce a formalism called \SMOK (scaled model of kinematics). 
		We define a goodness-of-fit statistic which is sensitive to
		deviations from uniform space motion, caused for example by binaries with periods of 10--50~years.
		 }
		{\HTPM will significantly improve the proper
motions of the \Hipparcos Catalogue well before highly accurate
\Gaia-only results become available. Also, \HTPM will allow us
to detect long period binary and exoplanetary
		candidates which would be impossible to detect from \Gaia data alone. The full
		sensitivity will not be reached with the first \Gaia release
		but with subsequent data releases. Therefore \HTPM should be repeated when more \Gaia
		data become available.}

\maketitle
%

\section{Introduction}
Stellar proper motions have traditionally been determined by analysing the
differences in position at different epochs, often separated by many decades
and obtained using vastly different instruments and methods. 
In this process, parallaxes (and radial motions, albeit relevant to a much
lesser extent) were mostly ignored.

With the advent of space astrometry, most notably the European satellite
\Hipparcos (1989--1993, see \citealt{hip:catalogue}), it became necessary
to treat data in a unified manner, i.e., by applying a single least-squares
solution for the position, parallax, and annual proper motion. 
\Hipparcos determined these parameters 
for nearly $120\,000$ stars\footnote{We use ``star'' to denote a catalogue
entry even when it refers to a non-single or extragalactic object. In 
the context of \Gaia data
processing the term ``source'' is commonly used for such objects. }
mostly brighter than magnitude $12$,
with a median uncertainty of about 1~milli-arcsecond (mas).
The \TychoTwo Catalogue \citep{2000A&A...355L..27H} gave additional data for 2.5 million stars observed with the \Hipparcos starmappers. The re-reduction of the \Hipparcos raw data \citep{book:newhip,fvl2007} significantly improved the main-mission results.
Today, 25~years after the launch of the satellite, these catalogues remain the main
source for the astrometric parameters of these stars.

The European space astrometry mission \Gaia will soon change this picture. 
\Gaia, launched at the end of 2013, will determine the astrometric parameters
of up to a billion stars between magnitude $6$ and $20$ with unprecedented
accuracies reaching a few tens of micro-arcseconds ($\mu$as) for \Gaia magnitude $G\la~15$. 
The vast amounts of data will be processed in a
single coherent least-squares solution, which solves not only for the
astrometric parameters but also for a large number of parameters describing the
time-varying spacecraft attitude and the geometry of the optical instrument. 
Due to the very large number of parameters to be determined from the
observational data the system cannot be solved directly
\citepads[][]{2010A&A...516A..77B} but has to be tackled in a block-iterative
manner with the so called ``Astrometric Global Iterative Solution'' (\AGIS).
The \AGIS software has been designed and implemented by groups at ESA/ESAC,
Lund Observatory and others, and is described in detail together with the
fundamental algorithms and mathematical framework by
\citetads{2012A&A...538A..78L}. 

Astrometric measurements obtained in the past, even of moderate accuracy by 
modern standards, have lasting value as they represent a state of the Universe
that is never repeated. A good example is the construction of proper motions in the \TychoTwo Catalogue using \Hipparcos and century-old photographic positions. When the astrometric parameters are propagated over a long
time interval, uncertainties in the tangential and radial motions accumulate to a significant 
positional uncertainty. Yet long-term deviations from linear space motion 
(e.g., in long-period binaries) increase even more drastically with time. Such deviations 
might not be detectable within the time spans of the \Hipparcos or \Gaia missions 
individually, but could be detectable by combining the results of the two.
Thus, although \Hipparcos will soon be superseded by \Gaia in terms of the 
expected accuracies at current epochs, its data form a unique comparison 
point in the past, very valuable in combination with later results.
For this reason the first \Gaia data release scheduled for 2016
will not only publish stellar positions and magnitudes based on the
first \Gaia observations, but also a combination of these observations with
the \Hipparcos Catalogue for all stars common between the two missions. This
part of the release is called the Hundred Thousand Proper Motions project
(\HTPM), originally proposed by F.~Mignard in a \Gaia-internal technical
document \citep[][]{LL:FM-040}.

This paper gives a recipe for the practical realisation of  the \HTPM project in the
context of the already existing \AGIS scheme for the astrometric solution of
\Gaia data. The proper motions in \HTPM might be trivially computed from 
the positional differences between an early \Gaia solution and the \Hipparcos
Catalogue -- the ``conventional catalogue combination'' approach of 
Sect.~\ref{sec:conventional}. However, we argue that the more elaborate 
``joint solution'' method described in Sect.~\ref{sec:jointsolution} will have
important advantages  
for the \HTPM project, and in Sect.~\ref{implementation} we show how to 
implement it as part of \AGIS. The validity and accuracy of the
method is demonstrated by means of a joint solution of
simulated \Gaia observations of the \Hipparcos stars (Sect.~\ref{sec:results}).
In the final sections we discuss the limitations of the results and their validity 
in the light of \Gaia's full nominal mission performance, as well as possible
applications of the joint solution method to other astrometric data.

The \HTPM project should use the re-reduction of the raw \Hipparcos data
\citep{fvl2007}, as it represents a significant improvement over the original
\Hipparcos Catalogue \citep{hip:catalogue}. Therefore it is also used in all our
simulations. For the purpose of demonstrating the \HTPM solution we regard all
valid entries of the \Hipparcos Catalogue as astrometrically well-behaved
(effectively single) stars. Their space motions are therefore regarded as uniform
(rectilinear, with constant speed) over the time interval covered by \Hipparcos
and \Gaia. This is obviously a very simplified picture of the true content of
the \Hipparcos Catalogue. However, getting the solution right in this simple
case is a first necessary step for any more sophisticated treatment of detected
binaries and multiple stars in the \Hipparcos Catalogue. 


\section{Theory} \label{theory}

Combining astrometric catalogues requires that data are expressed in the same
reference system and described in terms of a common kinematic model. 
In this section we describe the adopted model and how it is connected
to the definition of the astrometric parameters.
We outline the conventional approach to catalogue combination and develop
the ``joint solution'' as an optimal generalisation of the method. We show how to detect deviations
from the kinematic model or misfits between the datasets. We also outline how to
reconstruct the required information from \Hipparcos and how to integrate the
proposed scheme in the astrometric solution algorithm of \Gaia.

\subsection{Kinematic model of stellar motion}\label{sec:kinematicmodel}
The choice of astrometric parameters is a direct result of choosing a model of
stellar motion. 
The most basic assumption is for stars to move uniformly, i.e., linearly and
with constant speed, relative to the Solar System Barycentre (SSB). 
Note that this also means that the stars are assumed to be single.
This is obviously not true for all of them, but a good basic assumption for most stars. 
During the data reduction stars that are not ``well behaved'' in an astrometric
sense can be filtered out and treated further, e.g., by adding additional
parameters for components of stellar systems or for acceleration through
external influences.

A uniform space motion can be fully described by six parameters: 
three for the position in space at a chosen reference epoch, and three for the velocity. Traditionally, the
three positional parameters are right ascension $\alpha$, declination $\delta$,
and parallax $\varpi$ relative to the SSB at the reference epoch of the
catalogue.
The motion is then described by three parameters, where 
$\mu_{\alpha*}=\dot{\alpha}\cos\delta$ and $\mu_\delta=\dot{\delta}$ are the proper motions in right
ascension and declination, respectively, and the third parameter $\mu_r$ is 
the radial motion component.
The radial component is more commonly given as the radial velocity $v_r$ in
km~s$^{-1}$, but in an astrometric context it is conveniently expressed as the
radial proper motion 
(equivalent to the relative change in distance over time, or $-\dot{\varpi}/\varpi$)
\begin{equation}
\mu_r 
	= v_r \varpi / A \, ,
\label{eq:muR}
\end{equation}
where $A$ is the astronomical unit expressed in km~yr~s$^{-1}$. 
Only the first five parameters are classically considered astrometric
parameters. Based on only a few years of observations 
it is usually not possible to determine the
radial component from astrometry with sufficient accuracy \citepads{1999A&A...348.1040D}.
Hence the radial component is better determined by other techniques,
i.e., from spectroscopy. 
For \Gaia the radial component will be significant for many more stars,
although the affected fraction remains very small \citepads{2012A&A...546A..61D}.
Even though $\mu_r$ is not determined in the astrometric solution
for the vast majority of sources,
it is convenient and sometimes necessary to formulate astrometric problems 
with the full set of six astrometric parameters, as we do in this paper.
We will also show how to treat the sixth component when the radial velocity is
unknown or added from spectroscopy.

\subsection{Dealing with non-linearities: \SMOK}
When comparing and subsequently combining astrometric catalogues one needs to
deal with the fact that the mapping from rectilinear to spherical coordinates
is strongly non-linear. 
This becomes significant at the $\mu$as level when the differences in $\alpha$
and $\delta$ exceed some $(1~\mu\text{as})^{1/2}\simeq 0.5$~arcsec. 
For example, the barycentric direction traced out in $\alpha(t)$, $\delta(t)$
due to the proper motion will not be linear even though the star is assumed to
move uniformly through space. 
The traditional way to deal with this is to introduce higher-order correction
terms computed by Taylor expansion of the rigorous equations
\citep[e.g.,][]{taff1981}. 
In this paper we take a different approach, based on the ``scaled modelling of
kinematics'' (\SMOK) concept described in Appendix~\ref{sec:smok}. 
For the present purpose it is sufficient to know that $(\alpha,\,\delta)$ may
be replaced by linear coordinates $(a,\,d)$ relative to a designated, fixed
comparison point, with time derivatives $\dot{a}$, $\dot{d}$ representing the
components of proper motion in $\alpha$ and $\delta$. 
The six parameters $a$, $d$, $\varpi$, $\dot{a}$, $\dot{d}$, $\dot{r}$ (where
$\dot{r}$ is the \SMOK equivalent of the radial proper motion) provide an
alternative and equivalent parametrization of the kinematics, more convenient
for the catalogue combination than the usual set $\alpha$, $\delta$, $\varpi$,
$\mu_{\alpha*}$, $\mu_\delta$, $\mu_r$.


\subsection{Conventional catalogue combination}\label{sec:conventional}
In the conventional catalogue combination the astrometric parameters in each
catalogue are independently estimated from separate sets of 
observations, and the combination is done \aposteriori from the individual
catalogues.
Let ($a_1, d_1, \varpi_1$) at time $t_1$ be the position and parallax of a 
star in the first catalogue, and ($a_2, d_2, \varpi_2$) at time $t_2$ the
corresponding information in the second catalogue.  
The proper motion parameters $\dot{a}$, $\dot{d}$ are then derived as the positional
difference over time $\Delta t = t_1 - t_2$
\begin{align}	      
	\dot{a} 
		= (a_2 - a_1) 
		/ \Delta t \, ,
	\quad
	\dot{d} 
		= (d_2 - d_1) 
		/ \Delta t \, ,
\end{align}
which is possible thanks to the reformulation of the astrometric parameters in \SMOK. The proper motion uncertainties are 
\begin{align}
	 \sigma_{\dot{a}}
		= \frac{\sqrt{\sigma_{a1}^2 + \sigma_{a2}^2}}{\Delta t}\, , \quad 
	\sigma_{\dot{d}}
		= \frac{\sqrt{\sigma_{d1}^2 + \sigma_{d2}^2}}{\Delta t}\, ,\label{eq:propermotioncombination}
\end{align}
where $\sigma_{a1}$ is the uncertainty of $a_1$, etc.  
The third kinematic parameter $\dot{r}$ for the radial motion could in theory be derived from the
(negative, relative) difference in parallax, but in practice it is derived from the 
spectroscopic radial velocity as discussed in Sect.~\ref{sec:kinematicmodel}. 

While the proper motions are obtained by taking position differences over time, the
combined parameters for position and parallax are formed as 
weighted means. For $a$ this gives 
\begin{align}\label{eq:ahat}
	\hat{a} 
		= \frac{a_1\sigma_{a1}^{-2} 
		+ a_2\sigma_{a2}^{-2}}{\sigma_{a1}^{-2} 
		+ \sigma_{a2}^{-2}}\, ,
\end{align}
referring to the mean epoch of the combination
\begin{align}\label{eq:that}
	\hat{t}_a
		= \frac{t_1\sigma_{a1}^{-2} 
		+ t_2\sigma_{a2}^{-2}}{\sigma_{a1}^{-2} 
		+ \sigma_{a2}^{-2}} \, .
\end{align}
The reference time $\hat{t}_a$ is the optimal time in-between the two
catalogues at which the position and proper motion are uncorrelated and 
the uncertainty of $\hat{a}$ is minimal, given by 
	 $\sigma^{-2}_{\hat{a}} 
		= \sigma_{a1}^{-2} 
		+ \sigma_{a2}^{-2}$.
The expressions for $\hat{d}$ and $\hat{\varpi}$ are analogous.

This combination scheme has some limitations, in that it does not take
correlations between the astrometric parameters into account, nor the individual 
proper motions that may exist in each catalogue. In the next section we 
describe a more general approach.

\subsection{Joint solution}\label{sec:jointsolution}
The reduction of astrometric data is typically done using least-squares
solutions, resulting in a linear system of normal equations $\vec{N}\vec{x} =
\vec{b}$. Here, $\vec{x}$ is the vector of resulting astrometric parameters,
$\vec{N}$ the normal equations matrix, and $\vec{b}$ a vector constructed from
the residuals of the problem.%
\footnote{The least squares problem can be solved using a number of alternative 
numerical algorithms, for example based on orthogonal transformations. 
However, as these algorithms are all \emph{mathematically} equivalent to the
use of normal equations, our results remain valid independent of the chosen
solution algorithm.} 
The covariance $\vec{C}$ of the solution $\vec{\hat{x}}=\vec{N}^{-1}\vec{b}$ is formally given by
$\vec{C}=\vec{N}^{-1}$.

In \AGIS the observations of all well-behaved stars (``primary sources'') must be 
considered together in a single, very large least-squares solution (Sect.~\ref{sec:joint}). 
For $n$ primary sources, $\vec{x}$ would then be the full vector of $6n$ astrometric 
parameters, with $\vec{N}$ and $\vec{b}$ of corresponding dimensions. 
However, for the present exposition it is sufficient to
consider one star at a time, so that $\vec{x}$ and $\vec{b}$ are of length 6 
and $\vec{N}$ has dimensions $6\times 6$. In practice only five of the six
parameters are estimated, and $\vec{N}^{-1}$ should hereafter be regarded
as the inverse of the upper-left $5\times 5$
submatrix.\footnote{The full matrix is
nevertheless needed for the covariance propagation in Sect.~\ref{sec:rec}.}

On the assumption that the adopted kinematic model is valid for a particular
star, the matrix $\vec{N}$ and vector $\vec{b}$ encapsulate the 
essential information on the astrometric parameters, as determined by the
least-squares solution.
Thus, in order to make optimal use of the \Hipparcos data for a given star
there is no need to consider the individual observations of that star: 
all we need is contained in the ``information array'' $[\vec{N}~\vec{b}]$. In
Sect.~\ref{sec:rec} we show how this array is reconstructed from the published
\Hipparcos Catalogue.

Let $[\vec{N}_1~\vec{b}_1]$ and  $[\vec{N}_2~\vec{b}_2]$ be the information
arrays for the same star as given by two independent astrometric catalogues. 
From the way the normal equations are calculated from observational data it is
clear that the information arrays are additive, so that $[\vec{N}_1~\vec{b}_1]
+ [\vec{N}_2~\vec{b}_2]$ is the information array that would have resulted from
processing the two datasets together. 
In \citetads{2012ASPC..461..549M} we have proposed that the optimum combination of
the catalogues is done \apriori, that is by adding the corresponding arrays {\em before} solving. The result, 
\begin{align}\label{eq:js}
	\vec{\hat{x}} = (\vec{N}_1 + \vec{N}_2)^{-1} 
		(\vec{b}_1 + \vec{b}_2)  \, ,
\end{align}
is the {\em joint solution} of the astrometric
parameters, with covariance $\vec{\hat{C}} = (\vec{N}_1+\vec{N}_2)^{-1}$.
The two catalogue entries for the star must use the same reference epoch
and the same \SMOK comparison point.

The joint solution has several advantages over the conventional combination method outlined in
Sect.~\ref{sec:conventional}. 
Because it uses the full information in each catalogue it makes better use of
the data and allows to estimate the resulting uncertainties more accurately,
taking into account the correlations.
The individual proper motion information available in each catalogue is automatically
incorporated in the joint proper motion. 
Moreover, a solution might be possible where the data in each set individually
is insufficient to solve for all astrometric parameters, that is,
$\vec{N}_1+\vec{N}_2$ may be non-singular even if $\vec{N}_1$,
$\vec{N}_2$, or both, are singular. In practice if $\vec{N}_1$ comes from the
\Hipparcos data it will always be non-singular (since there is a \Hipparcos 
solution), and the sum is then also non-singular. Hence it will always be possible
to make a joint solution for all five astrometric parameters of the \HTPM stars, 
Finally, the joint solution scheme is a clean and rigorous approach and
can be integrated into the existing implementation of the astrometric solution
for \Gaia with moderate effort.

The joint solution can be seen as a multidimensional generalisation of the conventional scheme in
Sec.~\ref{sec:conventional}, with $\vec{N}$ representing the weights ($\sigma^{-2}$)
and $\vec{b}$ the astrometric parameters multiplied by their weights (e.g.,
$a\sigma^{-2}$). Then Eq.~(\ref{eq:js}) is the matrix equivalent of Eq.~(\ref{eq:ahat}).
The joint solution can also be understood in terms of Bayesian estimation
theory (assuming
multivariate Gaussian parameter errors), with $\boldsymbol{N}_1,
\boldsymbol{b}_{1}$ representing the prior information, $\boldsymbol{N}_2,
\boldsymbol{b}_{2}$ the new data, and their sums the posterior information. 

\subsection{Goodness of fit of the joint solution}\label{sec:gof}

The goodness of fit of a least-squares solution can be described in terms of
the sum of the squares of the normalized post-fit residuals,
\begin{equation}\label{eq:gof}
Q = \sum_k \left( \frac{\eta_k^{\rm (obs)} - \eta_k^{\rm (calc)}}{\sigma_k} \right)^2 \, ,
\end{equation}
where $\eta_k^{\rm (obs)}$ and $\eta_k^{\rm (calc)}$ are the observed and
calculated (fitted) angular focal-plane coordinates of the star in observation $k$, and $\sigma_k$
is the standard error of the observation. 
$Q$ is calculated for each star separately and is simply a function of
$\vec{x}=(a,\,d,\,\varpi,\,\dot{a},\,\dot{d},\,\dot{r})'$.  The least-squares
solution $\vec{\hat{x}}=\vec{N}^{-1}\vec{b}$ minimizes $Q$ and for any other
parameter vector $\vec{x}$ we have
\begin{equation}\label{eq:gof1}
Q(\vec{x}) = Q(\vec{\hat{x}}) + (\vec{x}-\vec{\hat{x}})'\vec{N}(\vec{x}-\vec{\hat{x}}) \, .
\end{equation}
If the kinematic model is correct and the standard errors of the observations are
correctly estimated one expects the minimum value $Q(\vec{\hat{x}})$ to
follow the chi-square distribution with $\nu$ degrees of freedom,
$Q(\vec{\hat{x}})\sim\chi^2(\nu)$. 
Here $\nu=m-\text{rank}(\vec{N})$ is equal to the number of observations $m$
(that is the number of terms in Eq.~\ref{eq:gof}) diminished by the rank of
$\vec{N}$. 
Note that this holds even if $\vec{N}$ is singular (i.e.,
$\text{rank}(\vec{N})<n$, where $n$ is the number of fitted parameters). 
In the singular case $\vec{\hat{x}}$ is not unique, yet $Q(\vec{\hat{x}})$ 
has a well-defined value (which may be 0 or positive). 

Analogous to Eq.~(\ref{eq:gof1}), in the joint solution we minimize the total
goodness of fit,
\begin{multline}\label{eq:gof2}
\quad Q(\vec{x}) = 
Q_1(\vec{x}) + Q_2(\vec{x}) =
Q_1(\vec{\hat{x}}_1) + Q_2(\vec{\hat{x}}_2) \\
+(\vec{x}-\vec{\hat{x}}_1)'\vec{N}_1(\vec{x}-\vec{\hat{x}}_1) +
(\vec{x}-\vec{\hat{x}}_2)'\vec{N}_2(\vec{x}-\vec{\hat{x}}_2) \, . 
\end{multline}
Here $\vec{\hat{x}}_i=\vec{N}_i^{-1}\vec{b}_i$ is the solution obtained by
using only catalogue $i=1,\,2$, i.e., minimizing $Q_i(\vec{x})$, which results in the minimum value $Q_i(\vec{\hat{x}}_i)$. 
It is readily seen that Eq.~(\ref{eq:gof2}) is minimized precisely for the
joint solution vector in Eq.~(\ref{eq:js}).

Each of the four terms in Eq.~(\ref{eq:gof2}) has a simple interpretation. 
The first term, $Q_1(\vec{\hat{x}}_1)$, is the chi-square obtained when fitting
the astrometric parameters only to the first set of data (in our case the
\Hipparcos data); similarly, $Q_2(\vec{\hat{x}}_2)$ is the chi-square obtained
when fitting only to the second set of data (from \Gaia).
The sum of the last two terms is minimized for $\vec{x}=\vec{\hat{x}}$, and
shows how much the chi-square is increased by forcing the \emph{same}
parameters to fit \emph{both} sets of data in the joint solution. 
This quantity is useful for assessing whether the two datasets are mutually
consistent and we therefore introduce a separate notation for it,
\begin{equation}\label{eq:deltagof}
\Delta Q = (\vec{\hat{x}}-\vec{\hat{x}}_1)'\vec{N}_1(\vec{\hat{x}}-\vec{\hat{x}}_1) +
(\vec{\hat{x}}-\vec{\hat{x}}_2)'\vec{N}_2(\vec{\hat{x}}-\vec{\hat{x}}_2) \, . \quad
\end{equation}
The two terms give the increase in chi-square due to the first and second dataset, respectively.

Long-period astrometric binaries may have significantly different proper
motions at the \Hipparcos and \Gaia epochs, and these in turn may differ from
the mean proper motion between the epochs. 
If the differences are significant, compared with the measurement precisions,
they will result in an increased value of $\Delta Q$. 
The null hypothesis, namely that the star is astrometrically well-behaved,
should be rejected if $\Delta Q$ exceeds a certain critical value. 
In order to calculate the critical value it is necessary to know the expected
distribution of $\Delta Q$ under the null hypothesis.
 
Let $m_i$ and $\nu_i=m_i-\text{rank}(\vec{N}_i)$ be the number of observations
and degrees of freedom in catalogue $i$. 
The number of degrees of freedom in the joint solution is
$\nu=(m_1+m_2)-\text{rank}(\vec{N}_1+\vec{N}_2)$. 
Under the null-hypothesis we have 
$Q_i(\vec{\hat{x}}_i)\sim\chi^2(\nu_i)$ ($i=1$, 2), $Q(\vec{\hat{x}})\sim\chi^2(\nu)$, 
and consequently 
\begin{equation}\label{eq:deltagofdistr}
\Delta Q \sim \chi^2(k) \, ,
\end{equation}
where
\begin{equation}\label{eq:deltagofdof}
k=\nu-\nu_1-\nu_2=\text{rank}(\vec{N}_1)+\text{rank}(\vec{N}_2)-\text{rank}(\vec{N}_1+\vec{N}_2) \, .
\end{equation}
In the special case when $\vec{N}_1$, $\vec{N}_2$, and $\vec{N}_1+\vec{N}_2$ all
have full rank (equal to $n$, the number of astrometric parameters) we have $k=n$.  
At a significance level of 1\% the critical values of $\Delta Q$, above which the null hypothesis
should be rejected, are 15.086, 13.277, 11.345, 9.210, and 6.635 for $k=5$, 4, 3, 2, and 1,
respectively \citep[e.g.,][]{abramowitz2012handbook}. 
With this criterion only 1\% of the well-behaved stars should be accidentally 
misclassified as not well-behaved. The expected distribution of $\Delta Q$ can be verified 
in the simulations which, by design, only includes well-behaved stars.

\subsection{Reconstruction of $\vec{N}_\textrm{H}$, $\vec{b}_\textrm{H}$ for the \Hipparcos Catalogue}\label{sec:rec}
When using the joint solution for incorporating \Hipparcos data in the
solution of early \Gaia data it is necessary to reconstruct the normal matrix
$\vec{N}_\textrm{H}$ and the right hand side $\vec{b}_\textrm{H}$ from
\Hipparcos for each star.
These are initially calculated for the reference epoch of the \Hipparcos
catalogue (J1991.25) and later propagated to the adopted reference epoch
of the joint solution (see Sect.~\ref{sec:joint}).

Let $a_\textrm{H}$, $d_\textrm{H}$, $\varpi_\textrm{H}$, $\dot{a}_\textrm{H}$,
$\dot{d}_\textrm{H}$ be the astrometric parameters from the \Hipparcos Catalogue
after transformation into the \SMOK notation (see Appendix~\ref{sec:smok}). 
 The upper-left $5\times 5$ submatrix of the covariance matrix can be taken
without changes from the \Hipparcos Catalogue (see Appendix~\ref{sec:hip2} for
details) since $\sigma_{\alpha*} = \sigma_a$, $\sigma_\delta = \sigma_d$,
$\ldots$ with sufficient accuracy at the reference epoch of the catalogue and
provided that the \SMOK comparison point is close enough to the astrometric
parameters of the star. 
The sixth parameter $\dot{r}_\textrm{H}$ and its corresponding entries in the
covariance matrix need to be added from external sources or set to sensible
values if not available (see below).  
Then the normal matrix is simply the inverse of the covariance matrix
$\vec{N}_\textrm{H} = \vec{C}_\textrm{H}^{-1}$ and
\begin{align}
	\vec{b}_\textrm{H} 
		= \vec{N}_\textrm{H} \begin{pmatrix}
			a_\textrm{H}, 
			d_\textrm{H},
			\varpi_\textrm{H},
			\dot{a}_\textrm{H},
			\dot{d}_\textrm{H},
			\dot{r}_\textrm{H}
		\end{pmatrix}'.
\end{align} 

\citet{hip:catalogue}{, Volume 1, Eq.~[1.5.69]} shows how to reconstruct the
elements $[\vec{C}_0]_{i6} = [\vec{C}_0]_{6i}$ $(i=1 \ldots 6)$, that is the sixth column and row of the
covariance matrix corresponding to the radial motion $\mu_r$. 
Let $\bar{v}_r$, $\bar{\varpi}$, $\bar{\mu}_r$ be the true values and $\delta v_r$,
$\delta \varpi$, $\delta \mu_r$ the errors. 
The expression in Eq.~[1.5.69] for the diagonal element $[\vec{C}_0]_{66}$ is
only valid if the relative
uncertainties in the radial velocity and parallax are small, i.e., $|\delta v_r
/ \bar{v}_r|$, $|\delta
\varpi / \bar{\varpi} | \ll 1$. 
If this is not the case we need to consider the complete expression for the calculated radial motion,
\begin{align}
\mu_r = \bar{\mu}_r + \delta \mu_r = (\bar{v}_r + \delta v_r) (\bar{\varpi} + \delta \varpi) / A\,,
\end{align}
where $\bar{\mu}_r = \bar{v}_r \bar{\varpi} / A$, leading to
\begin{align}
\delta \mu_r = (\bar{v}_r \delta \varpi + \bar{\varpi} \delta v_r + \delta v_r \delta \varpi) / A\,.
\end{align}
Squaring and taking the expectation while assuming that the errors in parallax and radial velocity are uncorrelated gives 
\begin{align}
\textrm{E}(\delta \mu_r^2) 
		= \left(\textrm{E}(v_r^2 \delta \varpi^2) 
		+ \textrm{E}(\varpi^2 \delta v_r^2) 
		+ \textrm{E}(\delta v_r^2 \delta \varpi^2)\right) / A^2\,,
\end{align}
where we replaced the true quantities by the observed ones.
The third term is the required generalization if $v_r$ or $\varpi$
is zero, or if the relative errors are large.
For example, if parallax and radial motion are unknown they could be assumed to be zero with a large
uncertainty.
The generalized version of Eq.~[1.5.69] in \citet{hip:catalogue} reads
\begin{align}
	[\vec{C}_0]_{66} 
		& = \left(v_{r0} / A\right)^2 [\vec{C}_0]_{33} 
		+ \left(\varpi_0 / A\right)^2 \sigma_{v{r0}}^{2} 
		+ \left(\sigma_{v{r0}} / A \right)^2 [\vec{C}_0]_{33},\nonumber\\
	[\vec{C}_0]_{i6} 
		& = [\vec{C}_0]_{6i} 
		= \left(v_{r0} / A \right)[\vec{C}_0]_{i3}, \quad i = 1 \ldots 5.
	\label{eq:sigmaVr}
\end{align}

The \Hipparcos Catalogue contains numerous entries for non-single stars,
for which additional parameters are given, describing deviations from uniform space motion. These additional parameters
are ignored in our simulations, which regard every star as single. In the
actual \HTPM solution many of these stars may require
more specialised off-line treatment. 
This is not further discussed in this paper.

\subsection{Joint solution in \AGIS}\label{sec:joint}
In reality the astrometric solution cannot be done separately for each star
as described in Sect.~\ref{sec:jointsolution}
but must consider all the stars together with the spacecraft attitude and
instrument calibration. Without prior information on the astrometric parameters this
leaves the solution undetermined with respect to the reference frame. 
This is not the case for the joint solution, however, as the \Hipparcos prior information
contains positions and proper motions that are expressed in a specific reference frame,
namely the \Hipparcos realisation of the International Celestial Reference System 
({\sc icrs}; \citeads[][]{1998A&A...331L..33F}).
The incorporation of the \Hipparcos prior in the joint solution automatically
ensures that the resulting data are on the \Hipparcos reference frame.
If required, the data can later be transformed into a more accurate representation
of the ICRS (see Sect.~\ref{sec:frame}). 

Due to the size of the data reduction problem \AGIS does not directly solve
$\vec{N}\vec{x}=\vec{b}$ but iteratively improves the astrometric parameters by
computing the updates $\Delta \vec{x}$, i.e., the difference to the current
best estimate values. When incorporating \Hipparcos data this requires us to
also express the \Hipparcos data (subscript H) as a difference to the current
best estimate (subscript c). Therefore we construct 
\begin{align}
	\Delta \vec{b}_\textrm{H} 
		= \vec{N}_\textrm{H}\Delta\vec{x} 
		= \vec{N}_\textrm{H} \begin{pmatrix}
			a_\textrm{H} - a_\textrm{c} \\
			d_\textrm{H} - d_\textrm{c}\\
			\varpi_\textrm{H} - \varpi_\textrm{c}\\
			\dot{a}_\textrm{H} - \dot{a}_\textrm{c}\\
			\dot{d}_\textrm{H} - \dot{d}_\textrm{c}\\
			\dot{r}_\textrm{H} - \dot{r}_\textrm{c}\\
		\end{pmatrix}.
\end{align} 
Before solving we add the corresponding matrices for the \Gaia data.  If no
additional \Gaia data would be added the solution would immediately
recover the \Hipparcos Catalogue parameters.

The reference epoch of the joint solution can be arbitrarily chosen. In
practice the \Gaia data are much better than the \Hipparcos data, therefore the
optimal reference epoch would always be very close to the epoch of the \Gaia
data alone. Assuming one releases \Gaia-only data and \HTPM results at the
same time it might be convenient to publish both for the same reference epoch,
i.e., the \Gaia-only reference epoch of the data release.


\section{Simulations}\label{implementation}

\subsection{Logic of simulations}
\begin{figure*}
\centering
	\usetikzlibrary{arrows,calc,positioning}
\usetikzlibrary{arrows,decorations.pathmorphing,backgrounds,fit,positioning,shapes.symbols,shapes,snakes,chains}
\begin{tikzpicture}[node distance=5em, auto, transform shape,scale=0.8,font=\sffamily]
	\tikzset{
		    mystyle/.style={draw=black, top color=white, minimum size=3em, text centered},
		    mynode/.style={mystyle, text width=10.5em, inner sep=1em, text centered},
		    mycatalogue/.style={mystyle, text width=10.5em, inner sep=1em, very thick, top color=black!10, bottom color=black!10, text centered, rounded corners},
		    myellipse/.style={ellipse, mystyle, text width=6em},
		    myarrow/.style={->, >=latex', shorten >=1pt, thick}, 
		    mylabel/.style={text width=7em} 
	}  

	\node[mynode] (hipparcos) {\Hipparcos catalogue\\(+ \XHIP/random $v_r$)}; 
	\node[mycatalogue, right=of hipparcos] (initial) 
		{\textbf{Initial catalogue}\\ (+ assumed $v_r$) \\Starting values for \AGIS 
		};
	\node[mycatalogue, right=of initial] (true) 
		{\textbf{Simulated ``true'' catalogue\\}
		};
	\node[mynode, right=of true] (background) {Auxiliary stars:\\Uniform sky, realistic distributions of $\varpi, \mu, v_r$};
	\node[myellipse, below=of true](observations) {Simulated Gaia observations};
	\node[myellipse, below=of observations](agis) {AGIS (astrometric solution)};
	\node[myellipse, below=of background](eval) {Comparison\\for evaluation and plotting};
	\node[myellipse, dashed, below=of hipparcos](prior) {prior information (optional)};
	\node[mycatalogue, below=of eval] (final) 
		{\textbf{Final catalogue}\\ Best estimate\\astrometry 
		};

	\draw[myarrow] (background.190) to (true.-10);
	\draw[myarrow] (true.190) to 
			(initial.-10);
	\draw[myarrow] (hipparcos.10) to (initial.170);
	\draw[myarrow] (initial.10) to 
			(true.170);
	\draw[myarrow] (initial.south) |- (agis.170);
	\draw[myarrow,-,dashed] (hipparcos.south) to (prior);
	\draw[myarrow,-,dashed] (prior.south) |- (agis.190);
	\draw[myarrow] (true) to (observations);
	\draw[myarrow] (observations) to (agis);
	\draw[myarrow,dashed] (final) to (eval);
	\draw[myarrow] (agis) to (final.173);

	\draw[myarrow,dashed] (true.330) -- ($(true.261)!.3!(eval.180)$) -| (eval);
\end{tikzpicture} 
    \caption{Relationships between catalogues during simulation runs.}\label{michalik_fig:catalogues}
\end{figure*}
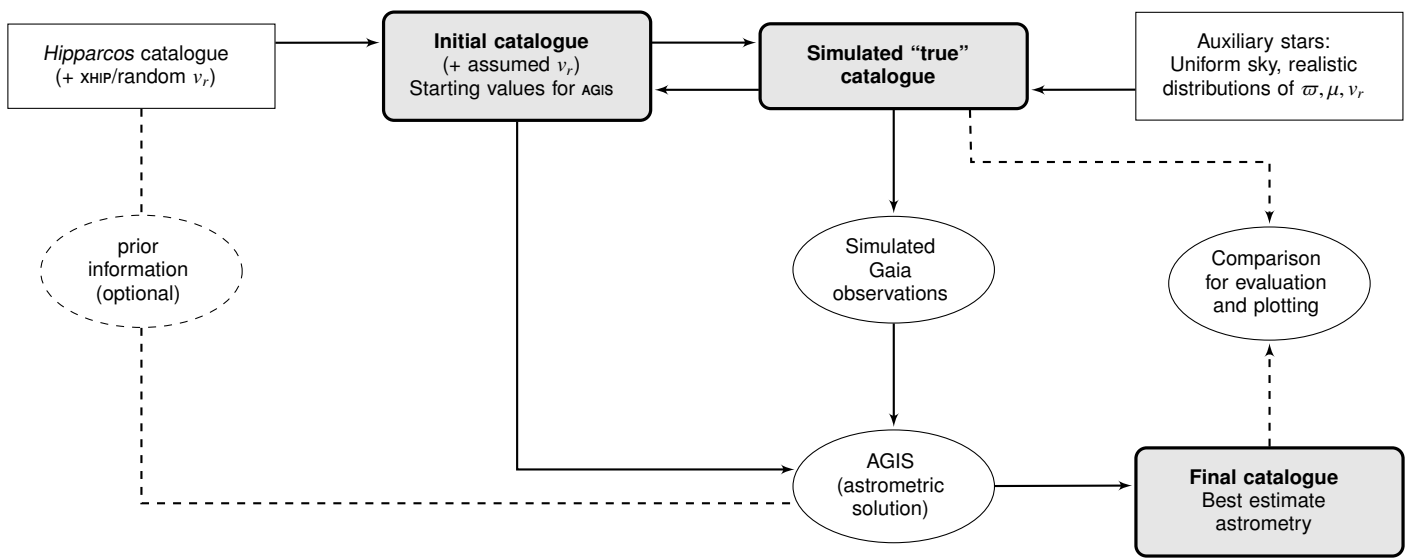
Simulations are based on \AGISLab \citepads{2012A&A...543A..15H}, a
small-scale version of the \AGIS data reduction created and maintained at Lund
Observatory. It is used to aid the development of algorithms for the astrometric data
reduction of \Gaia. Simulation runs are carried out in the following steps (cf.\
Fig.~\ref{michalik_fig:catalogues}):
\begin{enumerate}
	\item Creating catalogues of all the stars used in the simulation, namely the \Hipparcos 
	stars and the auxiliary stars (see below). Two catalogues are needed: 
	a simulated ``true'' catalogue to generate \Gaia observations and to evaluate the uncertainties of the astrometric performance, and an initial catalogue of starting values for the data reduction;\label{enum:truecatalogue}
	\item Simulating observations of the stars using the Nominal Scanning Law \citepads{2010IAUS..261..331D}, including perturbations 
	according to the expected precision of \Gaia measurements;
	\item Improving the astrometry of the initial catalogue through the astrometric solution (\AGIS),
	resulting in the final catalogue\label{enum:finalcatalogue}\label{enum:agis}. This can be done with or without incorporation of prior information from \Hipparcos;
	\item Evaluating the error of the resulting solution by comparing the final catalogue with the true catalogue.\label{enum:eval}
\end{enumerate}
Details of the first two steps are given below, while remaining steps are covered in Sect.~\ref{sec:results}. 

\subsection{Simulating the stellar catalogues}
All catalogues consist of two parts, the \Hipparcos stars and the additional
auxiliary stars. The \Hipparcos stars are necessary for the realisation of the
\HTPM scheme, and 113\,396 stars are within the nominal magnitude range of \Gaia ($G\simeq 6$--20). 
In order to obtain a reliable astrometric solution with a realistic modelling of the attitude
constraints we find that a minimum of one million stars is needed, uniformly
distributed on the sky. 886\,604 auxiliary stars are therefore added to the \Hipparcos stars in the solution.
The astrometric results for the auxiliary stars are not included in the statistics for the \HTPM performance, 
which is based only on the results for the \Hipparcos stars. However they contribute 
indirectly to the \HTPM solution via the attitude.


\subsubsection{Simulated ``true'' catalogue}\label{sec:simTrue}

The true catalogue defines the stars used for creating the simulated \Gaia
observations. For the real mission the true catalogue is of course not known.

To derive the \Hipparcos portion of the true catalogue we assume that the true
parameters deviate from the \Hipparcos values by random amounts consistent
with the \Hipparcos covariances. The \Hipparcos Catalogue is taken from CDS
and contains the astrometric parameters for the reference epoch J1991.25, 
including their covariance matrices (Appendix~\ref{sec:hip2}).
For each star let $\vec{C}$ be its covariance matrix, $\vec{L}$ the lower
triangular matrix resulting from the Cholesky decomposition $\boldsymbol{C} =
\boldsymbol{L}\boldsymbol{L}'$, and $\vec{g}$ a vector of six independent
standard Gaussian random variables (zero mean, unit standard deviation). Then the true parameters
(subscript T) are obtained by applying the error vector $\boldsymbol{e} =
\boldsymbol{L} \boldsymbol{g}$ to the astrometric parameters from the
\Hipparcos Catalogue (subscript H):
\begin{align}
	\vec{x}_\textrm{T} &= 
		\vec{x}_\textrm{H}+ \vec{e}.\label{eq:trueAstrometry}
\end{align}
Since $\text{E}(\boldsymbol{g}) = \boldsymbol{0}$, where $\text{E}(\ldots)$
denotes the expectation value, it follows that $\text{E}(\vec{e})=\vec{0}$. Moreover,
since $\text{E}(\boldsymbol{g}\boldsymbol{g}') = \boldsymbol{I}$ (the identity matrix), it
is readily verified that $\boldsymbol{e}$ has the desired covariance
$\text{E}(\vec{e}\vec{e}')=\vec{C}$. 
For a joint solution with simulated \Gaia data the \Hipparcos Catalogue needs to
be propagated to the reference epoch used in the solution. 

Rigorous propagation of the astrometric parameters must take into account
the radial motions of the stars, for which radial velocities are needed. We use data
from \XHIP \citepads{2012AstL...38..331A}, a compilation of radial
velocities and other data for the \Hipparcos stars from 47 different sources. We only
use radial
velocities with quality flag ``A'' or ``B'' in \XHIP. This makes for a total of
40\,171 radial velocities which are used as true values in our simulations.
For the remaining \Hipparcos stars we assign random radial velocities from a
Gaussian distribution with $v_r = 0$, $\sigma_{vr} = 30$~km~s$^{-1}$ using
Eq.~(\ref{eq:trueAstrometry}), based on the assumption that radial velocities
are typically smaller than that. 
The radial velocity uncertainty (taken from \XHIP or using 30~km~s$^{-1}$)
is also used to expand the $5\times 5$ covariance matrix by a sixth
column and row for the uncertainty and correlation of the radial motion, using
Eq.~(\ref{eq:sigmaVr}). 

For the auxiliary stars, the positions are chosen to give a random uniform
distribution across the sky with a mean density of about 21~stars deg$^{-2}$,
corresponding to one million stars needed for the solution.  We assume
magnitude $G=13$ for all auxiliary stars.  Since the number density of actual
stars with $G\le 13$ is about 60~deg$^{-2}$ at the Galactic poles, the assumed
distribution is a rather conservative estimate of the density of bright stars available for
the astrometric solution. The parallaxes
of the auxiliary stars are assumed to have a log-normal distribution with
median parallax 2.5~mas and a standard deviation of 0.6~dex.\footnote{Neglecting extinction, this 
corresponds to a Gaussian distribution of absolute magnitudes $M_G$ with 
mean value $+5$ and standard deviation 3~mag. This is not unreasonable for 
a local magnitude-limited stellar sample; cf.\ the HR diagram for nearby 
\Hipparcos stars, such as Fig.~1 in \citetads{1998MNRAS.298..387D}. 
The assumed distribution of true parallaxes and proper motions has some impact on our case B simulation results as discussed in Sect.~\ref{sec:2vs5}.}
The true proper motions and
radial velocities are calculated by assuming an isotropic velocity distribution
relative to the Sun with a standard deviation of 30~km~s$^{-1}$. 

\subsubsection{Initial catalogue and astrometric solution}
The initial catalogue contains the starting values for the data processing.
The \Hipparcos portion of it is identical to the astrometric parameters read from
the \Hipparcos Catalogue. For the auxiliary stars the initial positions are obtained by perturbing the true positions
with Gaussian noise of standard deviation 100~mas in each coordinate, while the initial
parallax and proper motion are set to zero. This is similar to a
real life scenario where one would assume initial stellar positions from
ground based observations or the first published \Gaia positions without
additional knowledge on the parallax or proper motion. The astrometric values in
the initial catalogue are subsequently updated by the \AGIS processing, resulting in
the final catalogue once the solution is found. We do not solve for the radial
motion but set the radial velocity to either zero (assuming no knowledge about it) or the true value (assuming it is perfectly known). 
In the first case perspective acceleration may show up for some stars as discrepancies 
in the solution, which disappear when the true radial velocities are used instead (see Sect.~\ref{discussion:perspectiveAcceleration}). 


\subsubsection{Final catalogue}
The final catalogue contains the astrometric parameters after data processing.
The difference to the simulated true catalogue gives the final errors of
the reduced data and is used to evaluate the quality of the astrometric results.
In this evaluation we focus on the improvement in the astrometric parameters 
of the \Hipparcos stars.

\subsection{Simulating \Gaia observations}
The observations of the one million stars described above are simulated using the 
Nominal Scanning Law of \Gaia. We neglect so called ``dead
time'' (when no data can be accumulated for example due to orbit
maintenance manoeuvres and
micro-meteoroid hits), which may amount to up to 15\% of the mission time.
We do however account for the dead time originating from stellar transits
coinciding with gaps between the CCDs in the focal plane, i.e., our simulations
remove such observations before further processing of the data.

To account for observation noise, i.e., the expected centroiding performance of
\Gaia, we use a simplified noise model that ignores the gating scheme that \Gaia
exploits for bright star detection. This noise model assumes a constant
centroiding performance for all \Hipparcos stars, identical 
to the centroiding performance for the brightest ungated stars at magnitude 13.
The typical along-scan standard error due to photon statistics is 94~$\mu$as.
A second noise component is added to account for various effects,
such as attitude modelling
errors \citepads{2013A&A...551A..19R} and uncertainties originating from
geometrical calibration parameters of the spacecraft. Although this additional
noise component may be correlated between individual CCD observations, we model it by quadratically adding a conservative RMS value of $300~\mu$as to the photon statistical standard error per CCD.

Based on the current \Gaia data release scenario\footnote{See \url{http://www.cosmos.esa.int/web/gaia/release} (2014 July 23). The first release of \Gaia data is foreseen for summer 2016. Discounting in-orbit commissioning, ecliptic pole scanning, and time for data processing leaves us
with about one year of \Gaia data.} we assume that the \HTPM project will initially be based on one year
of \Gaia data. The simulation results presented in Sect.~{\ref{sec:results}
use one year of \Gaia observations centred around the adopted reference epoch J2015.0.
\section{Results}\label{sec:results}
\subsection{Astrometric solution scenarios}
\begin{table}
\small
\centering
\caption{Number of astrometric parameters per star estimated in the four astrometric solution scenarios.
\label{tab:scenarios}}
\begin{tabular}{lcccc}
\toprule
  		&  \multicolumn{2}{c}{Case A (optimistic)}	& \multicolumn{2}{c}{Case B (conservative)}\\
\cmidrule[0.2pt](lr){2-3}
\cmidrule[0.2pt](lr){4-5}
		&  \G{12} & \HTPM			& \G{12} & \HTPM\\
\midrule
Hipparcos stars & 5	& 5			& 2 	& 	5\\
Auxiliary stars	& 5	& 5			& 2	&	2\\
\bottomrule
\end{tabular}
\end{table}

Table~\ref{tab:scenarios} gives an overview of the four 
different solution scenarios investigated in this paper.
The two cases called \G{12} do not use any prior data from the \Hipparcos Catalogue, but only
the 12~months of \Gaia observations. The other two, called \HTPM, use the \Hipparcos
covariances and astrometric parameters as priors in the processing of the same
\Gaia observations as in \G{12}. A comparison between the \HTPM and \G{12} scenarios thus 
allows to assess the improvement brought by the \Hipparcos prior information.

The scenarios are subdivided into cases A and B. In case A we
assume that there is sufficient \Gaia data to perform a full five-parameter
astrometric solution for all stars even without the \Hipparcos prior. This is an optimistic assumption, 
since in reality one year of data is only barely sufficient for a five-parameter solution under ideal
conditions, i.e., without data gaps. Dead time as outlined
before and the actual temporal distribution of observations over the year could mean
that the solution must be constrained to estimate only the two positional parameters for most of the stars. We
simulate this in case B by conservatively assuming that all stars for which we do not include a prior will have a two-parameter solution.
In such a solution the parallaxes and proper motions are effectively assumed to be zero, which gives
a large additional error component in the estimated positions.%
\footnote{Forcing a two-parameter
solution in case B for the stars without a prior creates residuals that are much larger than
the formal uncertainties of the \Gaia observations. The astrometric solution copes with this situation 
by means of the excess noise estimation described in Sect.~3.6 of 
\citetads{2012A&A...538A..78L}. Effectively this reduces the weight of the \Gaia observations
but does not affect the \Hipparcos prior. Without excess noise estimation the errors of the \HTPM proper motions
in case B  would be several times larger.\label{fn:excessnoise}}
While the \G{12}-B solution is then restricted to two parameters for all stars,
\HTPM-B can still solve all five parameters of the \Hipparcos stars.
Case B might be closer to the foreseen first release of \Gaia data and the first
release of \HTPM. Case A on the other hand demonstrates the capabilities of \Gaia and \HTPM
once sufficient data for a full astrometric solution are available in subsequent releases of \Gaia data.

\begin{table*}[t]
\small
\centering
\caption{Predicted uncertainties of the astrometric parameters of the \Hipparcos stars.
We compare the \Hipparcos data alone (\Hip) with a solution using only 12 months of \Gaia data (\G{12}), 
and a joint solution of \Hipparcos and \Gaia data (\HTPM). Case A and B refer to the optimistic and conservative
scenarios, respectively, described in the text. The two rightmost
columns give the predicted \HTPM proper motion uncertainties in the two cases.  \label{tab:results}}
\begin{tabular}{rrrrrrrrrrrrrrrrrr} 
\toprule
Mag.	& Number	& \multicolumn{6}{c}{Position [$\mu$as]} 	
			& \multicolumn{5}{c}{Parallax [$\mu$as]}	
			& \multicolumn{5}{c}{Proper motion [$\mu$as yr$^{-1}$]} \\
			\cmidrule[0.2pt](lr){3-8} 	
			\cmidrule[0.2pt](lr){9-13} 
			\cmidrule[0.2pt](lr){14-18}
	& 		&\Hip	& \Hip$_{2015}$		& \multicolumn{2}{c}{\G{12}}	& \multicolumn{2}{c}{\HTPM}	& \Hip	& \multicolumn{2}{c}{\G{12}}& \multicolumn{2}{c}{\HTPM}		& \Hip		& \multicolumn{2}{c}{\G{12}}	& \multicolumn{2}{c}{\HTPM}\\
			\cmidrule[0.2pt](lr){5-6} 	
			\cmidrule[0.2pt](lr){7-8} 	
			\cmidrule[0.2pt](lr){10-11} 	
			\cmidrule[0.2pt](lr){12-13} 	
			\cmidrule[0.2pt](lr){15-16} 	
			\cmidrule[0.2pt](lr){17-18} 	
	& 		&	& 		& A 	& B\tablefootmark{\phantom{a}}& A	& B	& 		& A	& B	& A	& B\tablefootmark{\phantom{b}}&		& A	& B	& A & B \\
\midrule[0.2pt]                                                                 
6--7	&  9\,381 	& 367	&  10\,892	& 41	&3\,388\tablefootmark{\phantom{a}}& 36	& 312	&    501	& 82	& -	& 43	& 250\tablefootmark{a}     &    458	& 207	& -	& 14& 27 \\
7--8	& 23\,679	& 497	&  14\,434	& 41	&2\,692\tablefootmark{\phantom{a}}& 35 	& 318	&    684	& 81	& -	& 43	& 261\tablefootmark{a}     &    608	& 204	& -	& 19& 30 \\
8--9	& 40\,729	& 682	& 19\,947	& 41	&2\,369\tablefootmark{\phantom{a}}& 35	& 330	&    939	& 77	& -	& 43	& 271\tablefootmark{a}     &    840	& 197	& -	& 26& 35 \\
9--10	& 27\,913	& 936	& 27\,629	& 40	&2\,663\tablefootmark{\phantom{a}}	& 35	& 333	& 1\,284	& 77	& -	& 43	& 274\tablefootmark{a}     & 1\,165	& 194	& -	& 35& 43 \\
10--11	&  8\,563	& 1\,403& 41\,352	& 42	&5\,240\tablefootmark{\phantom{a}} 	& 36	& 343	& 1\,921	& 83	& -	& 46	& 283\tablefootmark{a}     & 1\,744	& 205	& -	& 50& 60 \\
11--12	&  2\,501	& 2\,125& 61\,896	& 41	&13\,687\tablefootmark{\phantom{a}}	& 35	& 357	& 2\,882	& 78	& -	& 47	& 291\tablefootmark{a}     & 2\,607	& 195	& -	& 70& 85 \\
$\ge$12	&     630	& 3\,248&109\,030	& 42	&13\,926\tablefootmark{\phantom{a}}	& 38	& 378	& 4\,291	& 80	& -	& 51	& 295\tablefootmark{a}     & 4\,578	& 209	& -	& 94& 134 \\
\midrule[0.2pt]
all	& 113\,396	& 753	& 22\,148	& 41 	& 2\,856\tablefootmark{\phantom{\phantom{a}}}& 35	& 328	& 1\,033	& 79	& -	& 44	& 271\tablefootmark{a}   &    932	& 199	& -	& 29& 38 \\
\bottomrule
\end{tabular}
\tablefoot{\tablefoottext{a}{Case B parallaxes are biased as shown in Fig.~\ref{fig:varpiHist}. This bias is not included in the RSE values given here.}}
\end{table*}

\subsection{Predicted astrometric accuracies of \HTPM}
Table~\ref{tab:results} summarizes the results for the entire set of \Hipparcos stars, and subdivided by magnitude.
No results are given for the auxiliary stars, but they are similar to the results for the \Hipparcos stars in the \G{12} scenarios.
For comparison we also give the formal uncertainties from the \Hipparcos Catalogue. 
For the positions they are given both at the original epoch J1991.25 and at the
epoch J2015 of the \Gaia data. It should be noted that the simulations
include stars which in the \Hipparcos Catalogue are described
with more than five parameters, but are here treated as single stars. Excluding
them from the statistics would systematically reduce the \Hipparcos
uncertainties in Table~\ref{tab:results}. The real \HTPM solution will also
include all \Hipparcos
stars independent of the type of solution in the \Hipparcos Catalogue. A poor
fit between the \Gaia and \Hipparcos data will then be used to filter out
binary candidates for further treatment.

All \G{12} and \HTPM uncertainties in Table~\ref{tab:results} are
derived from the distribution of the actual errors (calculated values minus
true values) obtained in the solutions, using the ``Robust Scatter Estimate'' (RSE).%
\footnote{The RSE is
	defined as 0.390152 times the difference between the 90th and 10th percentiles
	of the distribution of the variable. For a Gaussian distribution it equals the
	standard deviation. Within the \Gaia core processing community the RSE is used as a standardized, 
	robust measure of dispersion \citepads{2012A&A...538A..78L}.\label{footnote:RSE}}
Rather than stating the uncertainty of $\alpha$ and $\delta$ separately we
give the mean of the RSE in the two coordinates as the position uncertainty.
Similarly the proper motion uncertainty is the mean RSE of the errors in
$\mu_{\alpha *}$ and $\mu_{\delta}$. 

\paragraph{Proper motion}
The joint solution shows a big improvement in the proper motion 
uncertainties compared with the \Hipparcos data.  
The improvement factor of \HTPM compared with \Hipparcos alone is 32 in case A 
and 25 in case B. The factors are similar because the \Hipparcos position 
uncertainty dominates over the \Gaia uncertainty in both cases. 
In the optimistic case A, the proper motions from the \Gaia-only data are already
better than \Hipparcos alone, but not as good as the joint \HTPM solution. 

Using Eq.~(\ref{eq:propermotioncombination}) to estimate the expected
precision of the conventional combination we find in case~A proper motions of 16 and
137~$\mu$as~yr$^{-1}$ for the brightest and faintest magnitude bins, compared
with 14 and 94~$\mu$as~yr$^{-1}$ in the \HTPM-A results. In case~B
we find 143 and 602~$\mu$as~yr$^{-1}$, respectively,
compared with 27 and 134~$\mu$as~yr$^{-1}$ in \HTPM-B. The joint solution thus gives
consistently better results as discussed in Sect.~\ref{sec:jointsolution}.

\paragraph{Parallax}
The improved proper motions allows better to disentangle the five parameters 
in the joint astrometric solution (cf.\ Fig.~\ref{fig:varpiMap}), resulting in 
improved parallax uncertainties.
In case A we find that the parallax uncertainties in the joint solution improve by a factor
23 compared with \Hipparcos, and a factor 2 compared with \G{12}.
However in the more realistic case B
the improvement is much smaller (a factor 3 compared with \Hipparcos) and the parallaxes are
strongly biased as shown in Fig.~\ref{fig:varpiHist}. This bias originates from the assumption of zero
parallax and proper motion in the two-parameter solution of the auxiliary
stars. The true positive parallaxes result in a biased attitude, which
propagates into the five-parameter solution of the \Hipparcos stars making their
parallaxes systematically too small. (As discussed in Sect.~\ref{sec:2vs5}, this bias can
be entirely avoided in later releases of \Gaia data through a proper selection of primary sources.)

\paragraph{Position}
The extremely good \Gaia observations lead to an improvement by up to a 
factor $\sim\,$600 compared with \Hipparcos positions propagated to J2015. 
In case A the slight improvement in the \HTPM positions compared with \G{12}  
comes from the better determination of  proper motion and
parallax. In case B the \Gaia-only positions show a large uncertainty due to the
two-parameter solution which neglects the true parallaxes and proper motions of
the stars. The increase in position uncertainties is especially pronounced for the fainter 
stars due to preferential selection of nearby high-proper motion stars in the non-survey 
part of the \Hipparcos Catalogue, which means that their (neglected) parallaxes and
proper motions are statistically much larger than for the brighter (survey) stars.
In the \HTPM solution for case B all five parameters are solved for the \Hipparcos stars,
so the sizes of their parallaxes and proper motions have no direct impact on the 
accuracy of the solution. However, the positional uncertainties are still much
increased compared with case A, because the two-parameter solutions for the 
auxiliary stars degrade the attitude estimate. 

\begin{figure}
\centering
\includegraphics[width=0.4\textwidth]{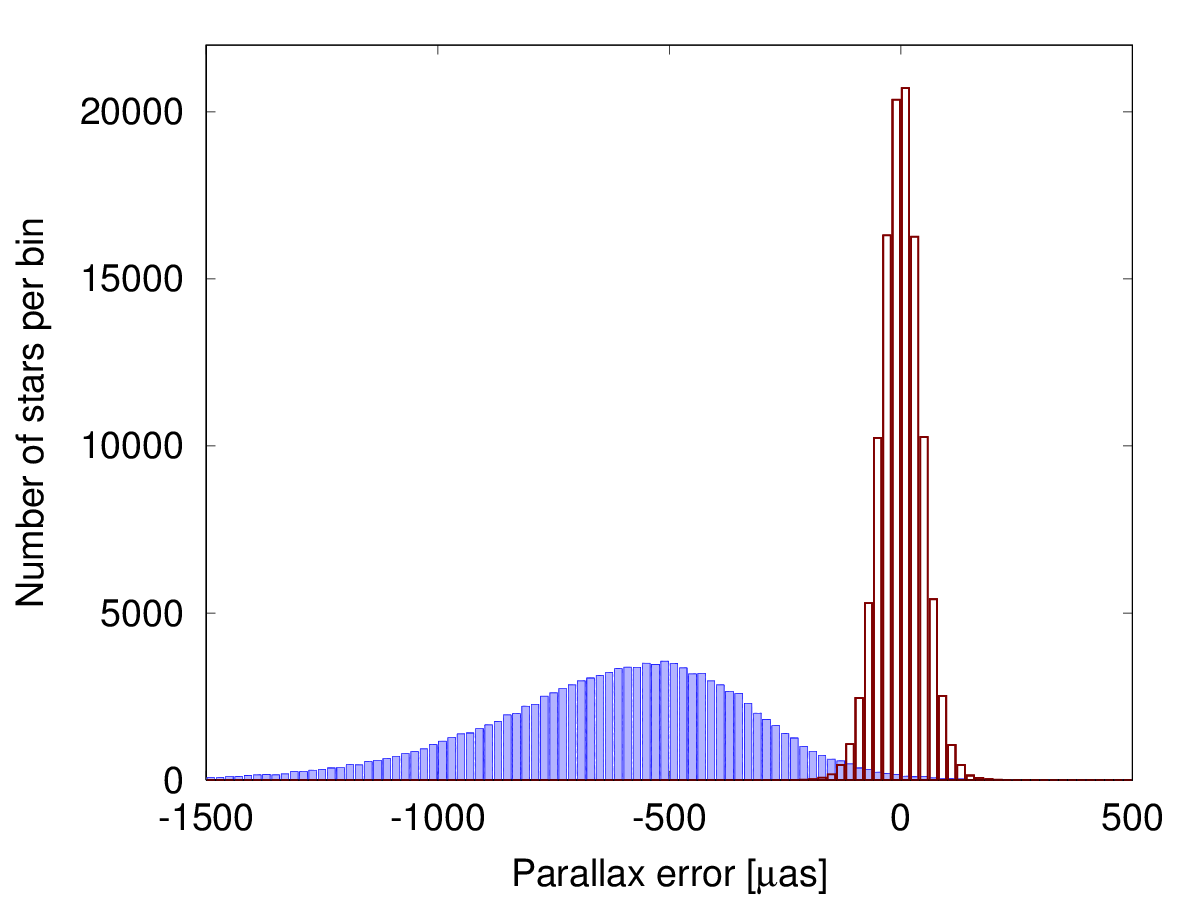}
\caption{Histograms of the parallax errors in the \HTPM solution for two cases. Bin width is 20~$\mu$as. In case A (full five-parameter astrometric solution for all stars, red/right histogram) the parallax errors are unbiased. In case B (two-parameter solution of the auxiliary stars, blue/left histogram) the median parallax error is $-591~\mu$as.\label{fig:varpiHist}}
\end{figure}

\begin{figure*}[]
\centering
\includegraphics[width=0.49\textwidth,clip,trim=0 0 0 0]{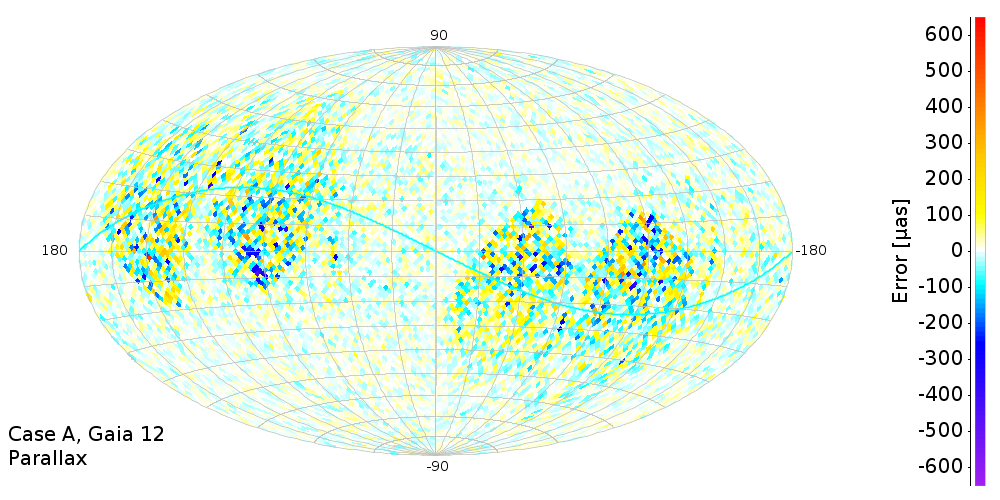}
\includegraphics[width=0.49\textwidth,clip,trim=0 0 0 0]{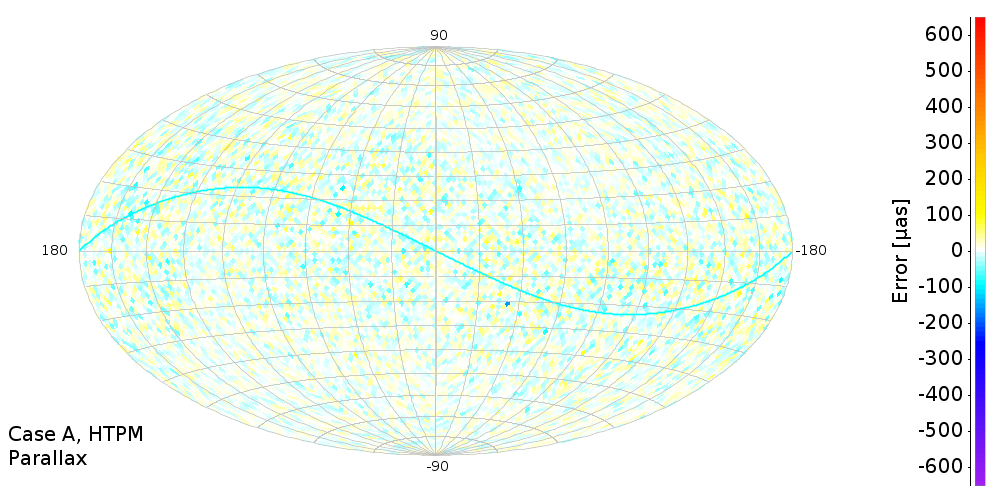}\\
\includegraphics[width=0.49\textwidth,clip,trim=0 0 0 0]{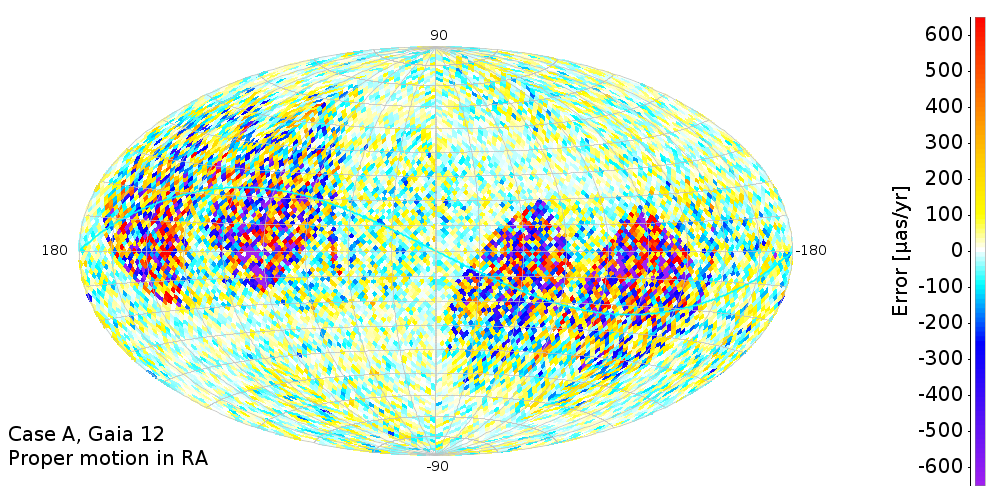}
\includegraphics[width=0.49\textwidth,clip,trim=0 0 0 0]{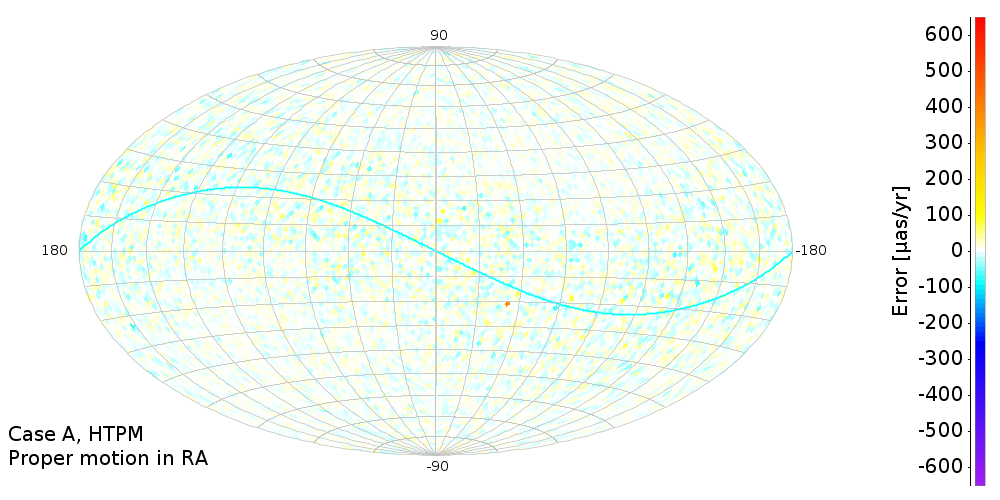}\\
\includegraphics[width=0.49\textwidth,clip,trim=0 0 0 0]{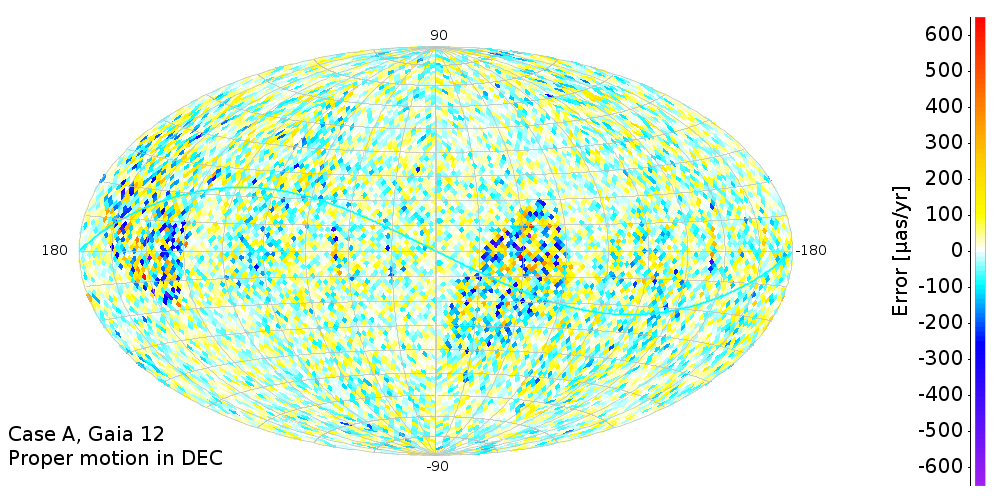}
\includegraphics[width=0.49\textwidth,clip,trim=0 0 0 0]{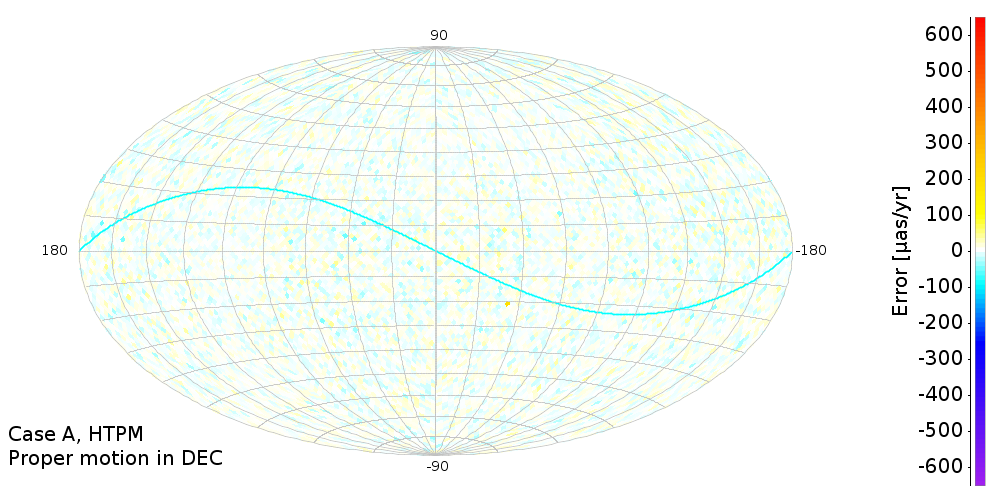}\\
\caption{Distribution of the parallax and proper motion errors on a Hammer-Aitoff equatorial
projection of the sky. All maps are for case A (full five-parameter solutions for all stars). 
The left figures show the results from the 12 months' \Gaia-only simulation. 
Some regions of the sky are poorly observed resulting in zonal errors. 
The right figures show the \HTPM results for the same stars.
The prior helps to disentangle proper motion and parallax, therefore we find a
more homogeneous distribution of errors at an overall lower level. The cyan line follows the ecliptic for reference.\label{fig:varpiMap}}
\end{figure*}
\subsection{Goodness of fit statistics\label{discussion:perspectiveAcceleration}}
\begin{figure*}
\centering
\includegraphics[width=0.4\textwidth]{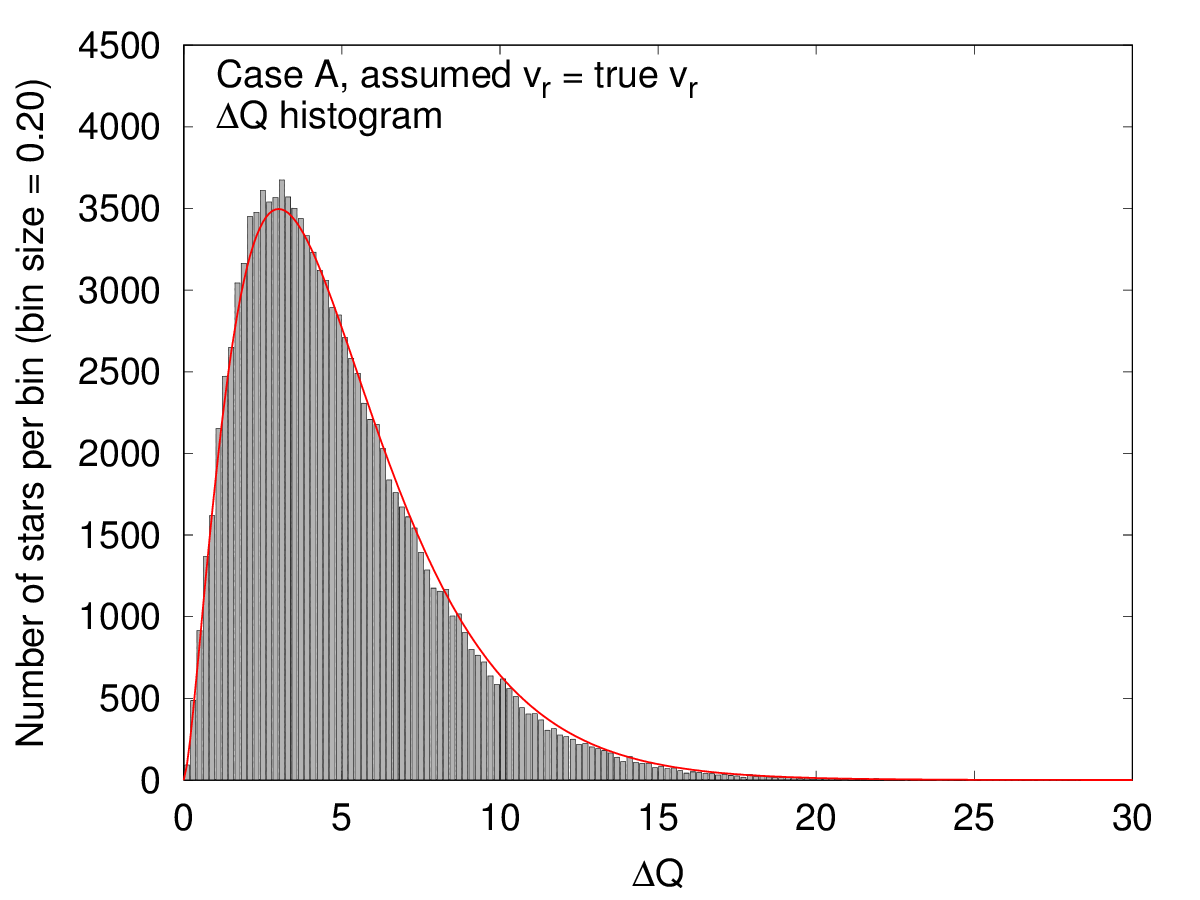}
\includegraphics[width=0.4\textwidth]{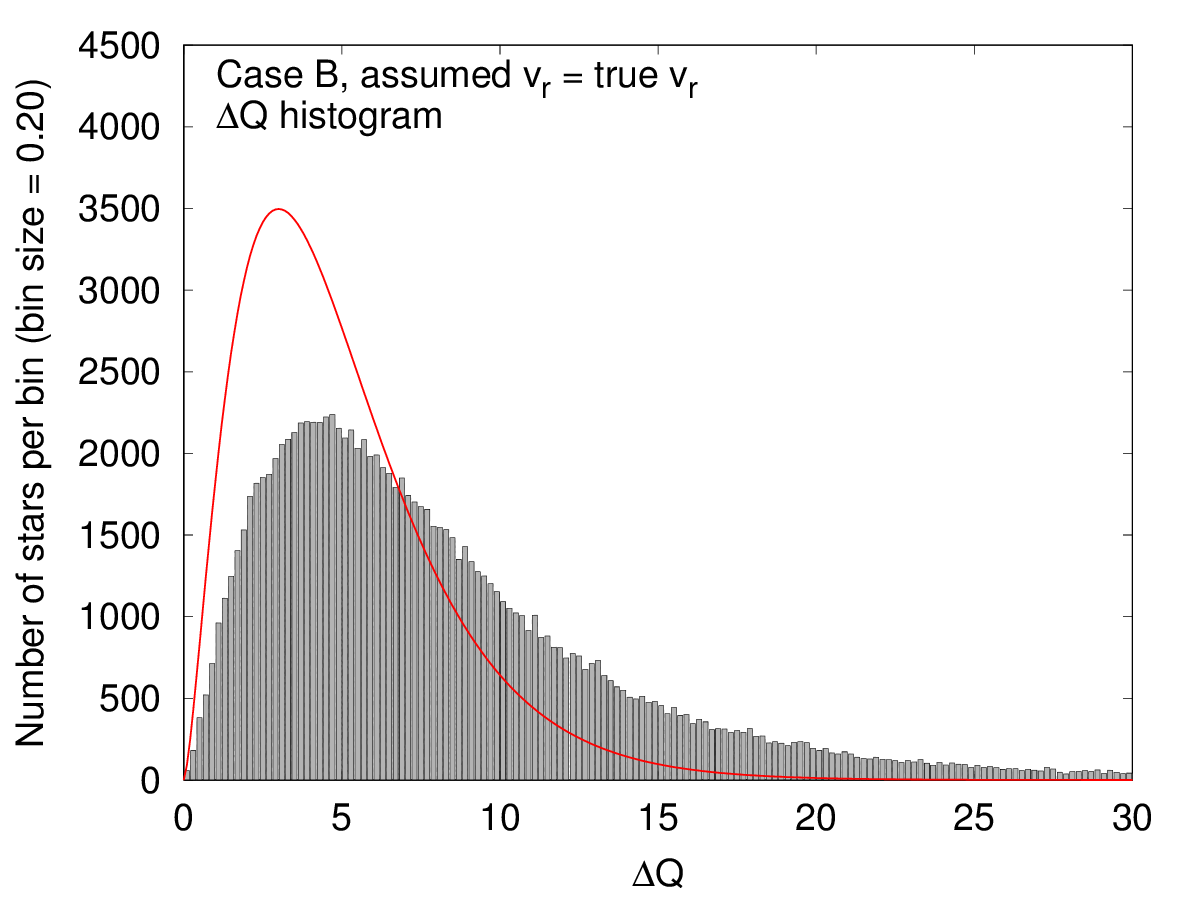}\\
\includegraphics[width=0.4\textwidth]{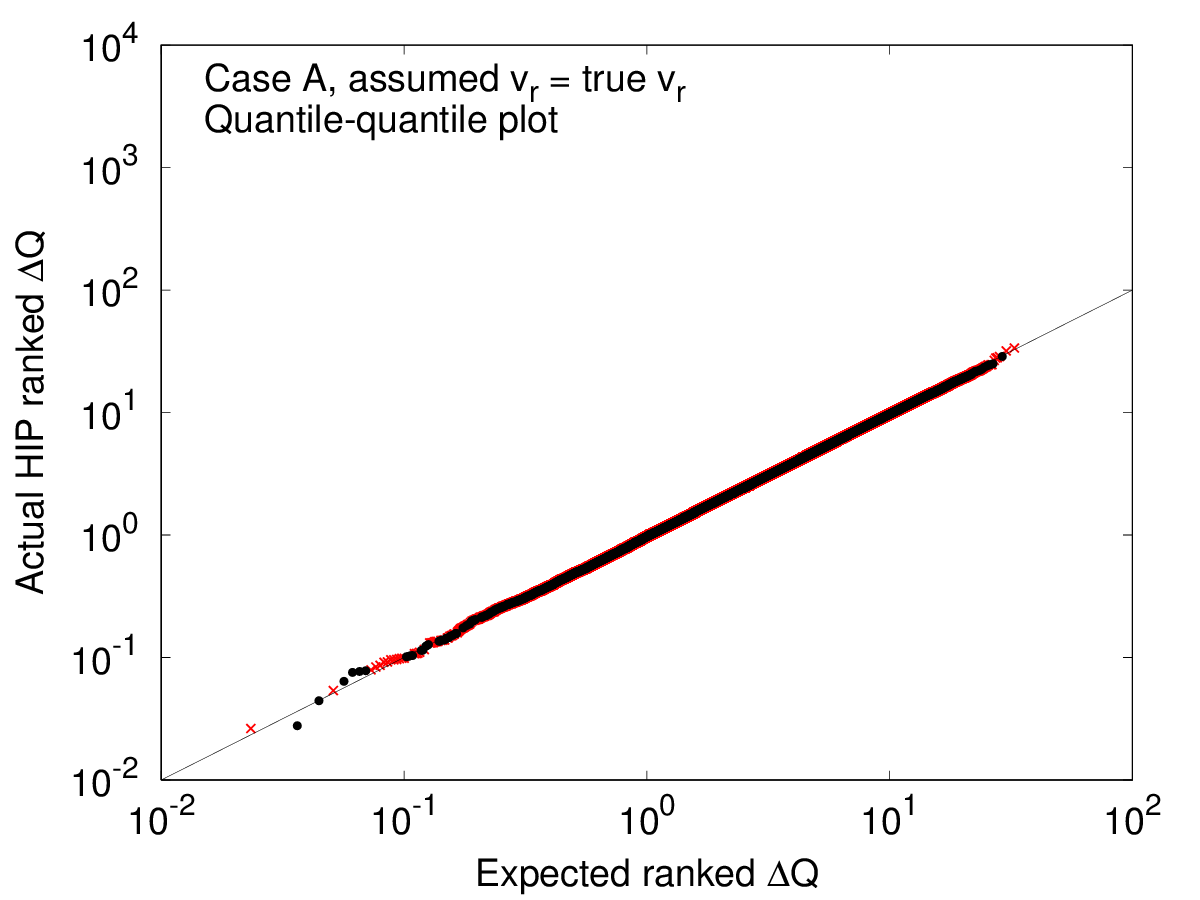}
\includegraphics[width=0.4\textwidth]{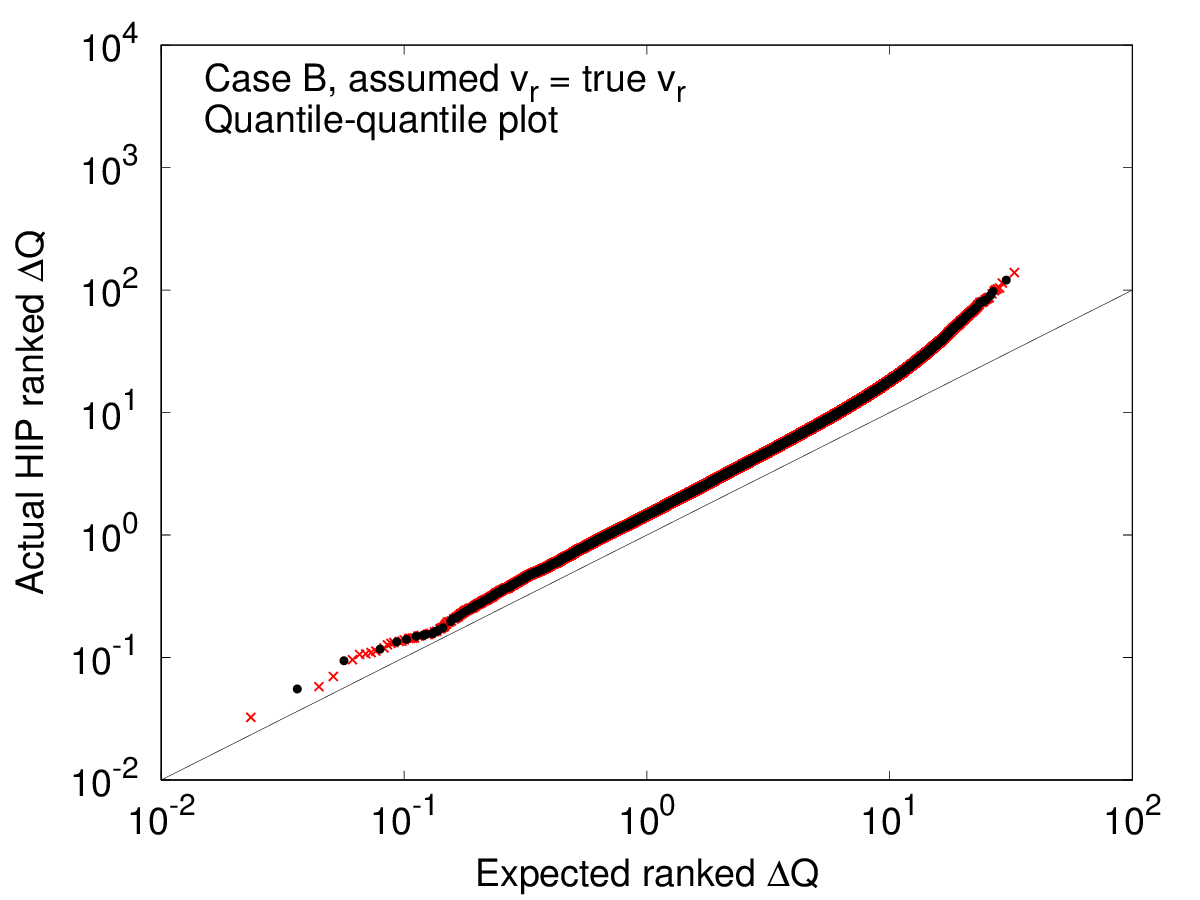}\\
\includegraphics[width=0.4\textwidth]{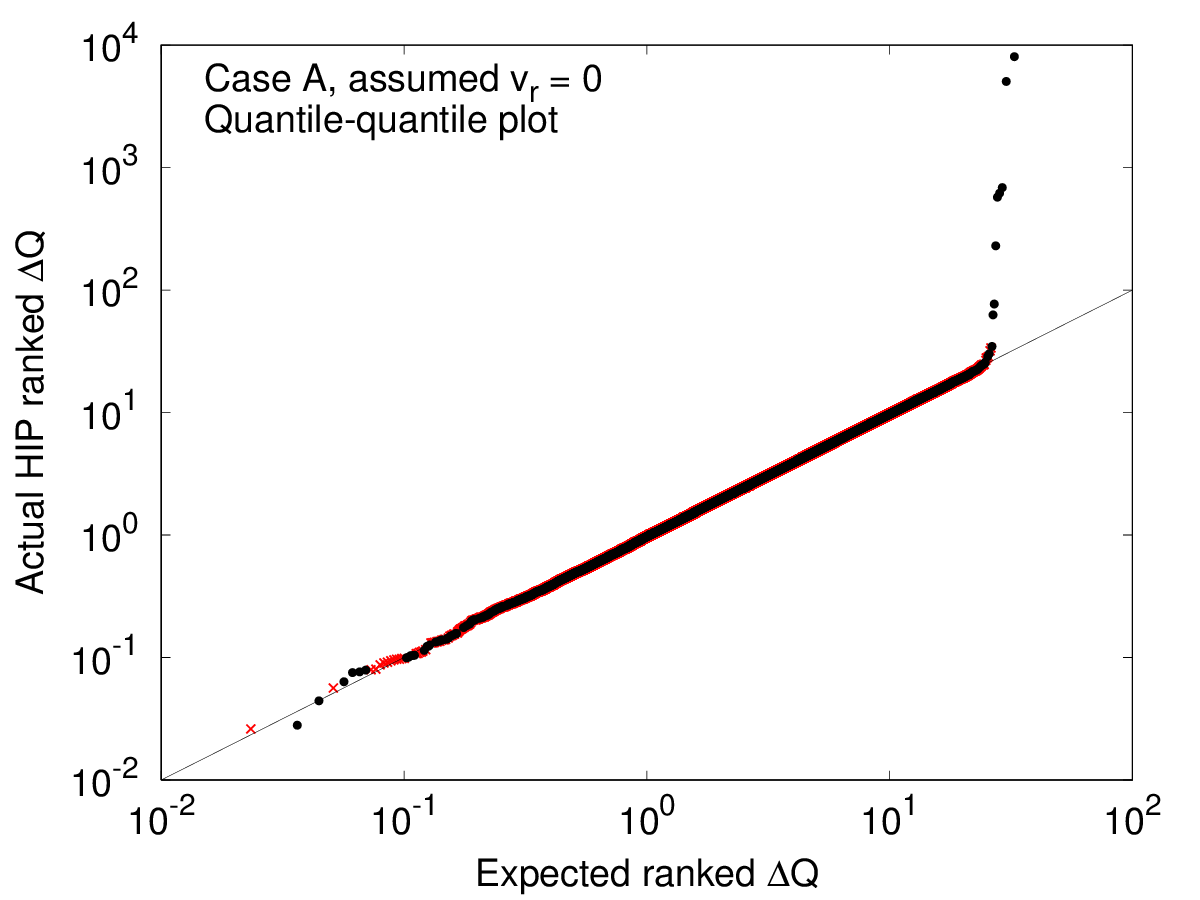}
\includegraphics[width=0.4\textwidth]{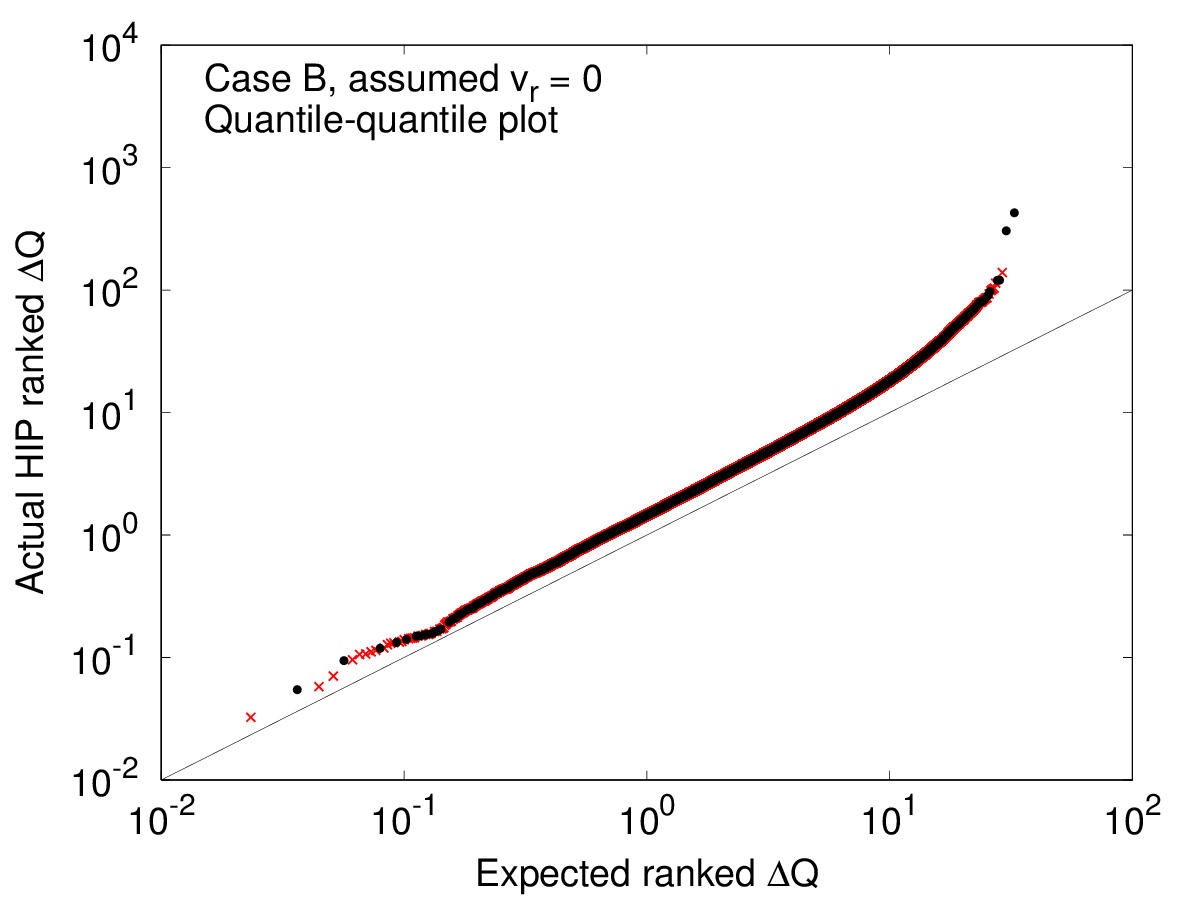}\\
\includegraphics[width=0.4\textwidth]{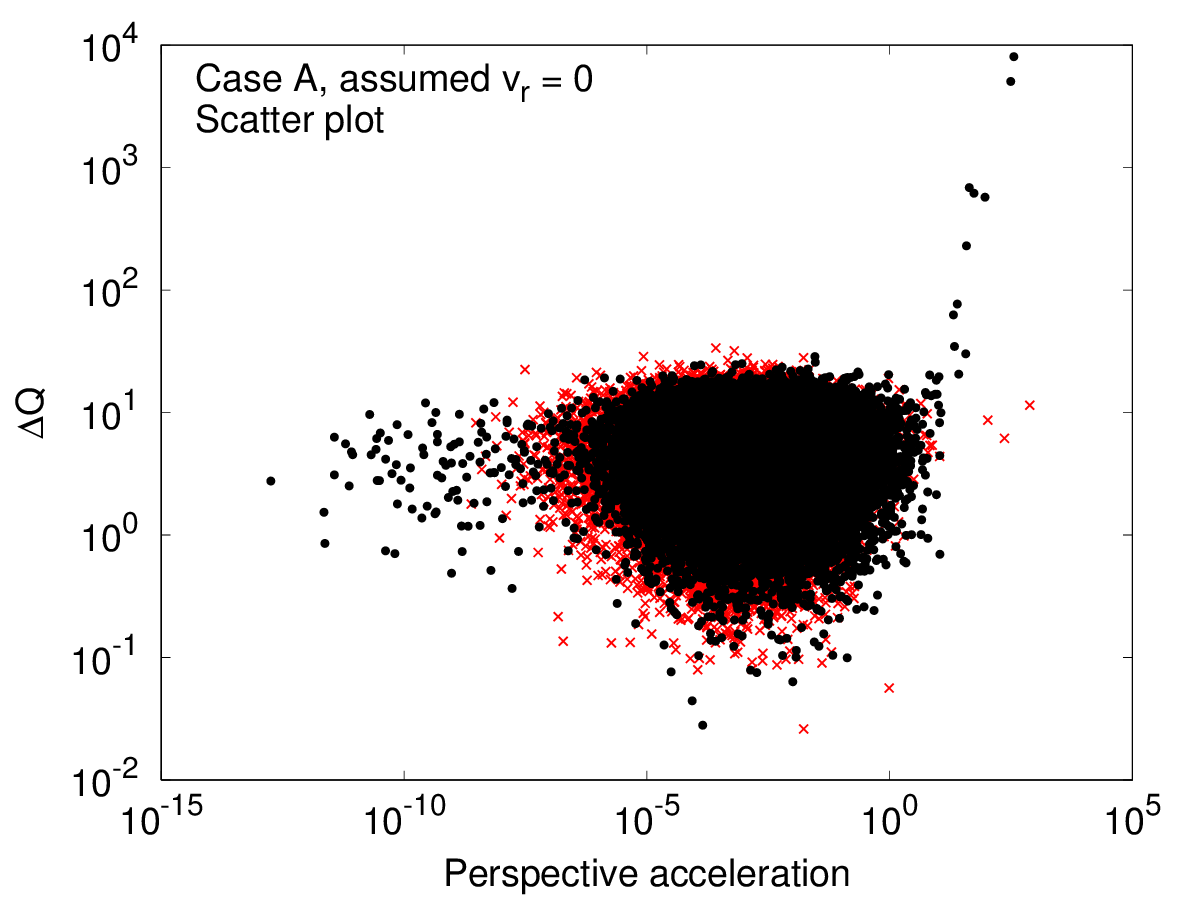}
\includegraphics[width=0.4\textwidth]{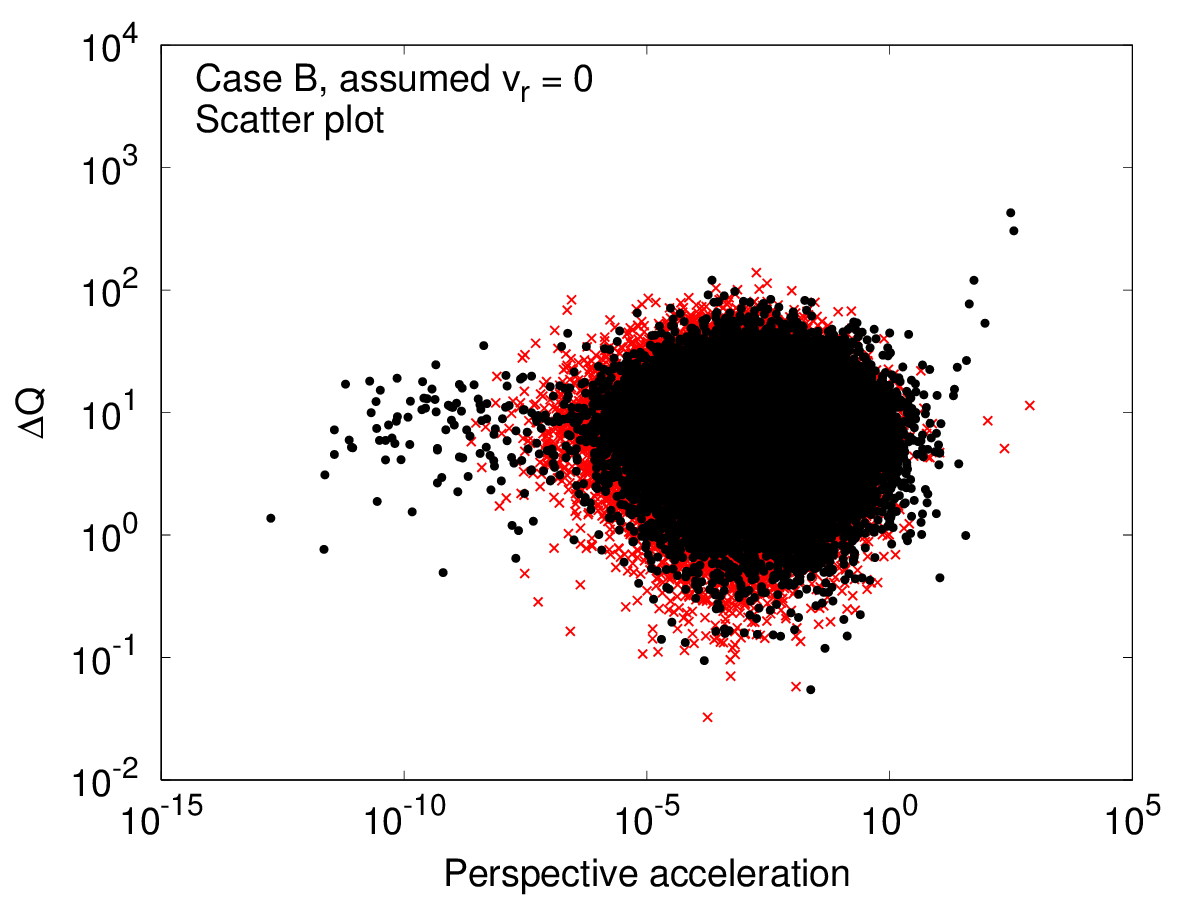}
\caption{Left column: Goodness of fit values $\Delta Q$ for case A simulations.
From top to bottom, the histogram of $\Delta Q$ values (grey bars) follows a $\chi^2$
distribution (red line) with five degrees of freedom. If the
assumed radial velocities in the solution equal the true values, the actual and expected distribution agree perfectly.
If the assumed radial velocity is unknown (set to zero) deviations
from the expected distribution are seen. These outliers are caused by
perspective acceleration. The markers in the quantile-quantile and scatter plots correspond to stars with
radial velocities from \XHIP (black dots) and to stars with random radial
velocities (red crosses). The three rightmost red crosses in the scatter plots correspond to
HIP80190, HIP80194 and HIP67694 which have very large uncertainties in the
\Hipparcos Catalogue. Therefore they do not show a large $\Delta Q$ value even
though they have large perspective acceleration.
The right column shows the same plots for case B simulations (see Sect.~\ref{sec:gof}).
\label{fig:deltaQ}}
\end{figure*}

As discussed in Sect.~\ref{sec:gof} the goodness of fit value $\Delta Q$ from 
Eq.~(\ref{eq:deltagof}) describes how well the joint astrometric solution fits the individual
observations of both missions together. If all the observations are consistent with the
kinematic model, then $\Delta Q$ is expected to follow a $\chi^2$ distribution with five 
degrees of freedom. Larger values indicate deviations from the model, for example
non-uniform motion caused by invisible companions or astrometric binaries. In the
present simulations we do not include any such objects, so we expect $\Delta Q$ to 
follow the theoretical distribution. 

The top two diagrams in the left column in Fig.~\ref{fig:deltaQ} shows that this is indeed true in case
A, if the radial velocities assumed in the solution are the true ones. The result would have 
been the same if the assumed radial velocities had only been wrong by a few km~s$^{-1}$.
If instead we assume zero radial velocities for all stars, as was done in the bottom two diagrams (while the observations were                                                                                           
still generated with non-zero radial velocities), we find a small number of
outliers. It turns out that all
of them are nearby, high-velocity stars (Table~\ref{tab:outlierlist}) expected to show
significant perspective acceleration, that is the change in proper motion due to the changing stellar
distance and the changing angle between the line of sight and motion of the
star (\citeads{1917AJ.....30..137S}; \citeads{1977VA.....21..289V}; \citealt{murray1983}). 
This perspective acceleration is not taken
into account in the solution when the radial velocities are assumed to be zero,
giving a mismatch between the \Hipparcos data and the observed \Gaia position.
The positional offset due to the perspective acceleration after $\Delta t$ years
amounts to
\begin{equation}\label{eq:persp}
\Delta\theta_\text{persp} = \mu\mu_r\Delta t^2 \, ,
\end{equation}
where $\mu=(\mu_{\alpha*}^2+\mu_\delta^2)^{1/2}$ is the total proper motion.
As shown in Table~\ref{tab:outlierlist}, the stars with a large $\Delta Q$ also have
a large offset $\Delta\theta_\text{persp}$ at the \Gaia epoch, compared with the
positional uncertainty of the solution at that epoch.

This demonstrates that knowledge of radial velocities is required for a number
of stars to avoid false positives in the detection of non-uniform space motion
\citepads{2012A&A...546A..61D}. It also shows that $\Delta Q$ is a useful statistic
for detecting non-uniform space motion in general.

The right column in Fig.~\ref{fig:deltaQ} shows the corresponding results in case B. Here $\Delta Q$
follows a scaled version of the expected distribution with a somewhat extended tail.
The two bottom panels show that $\Delta Q$ is still a useful measure of
deviations from the adopted kinematic model although it is much less sensitive
than in case A. As a result only two outliers due to the perspective
acceleration are found if the assumed radial velocities are set to zero. 
This demonstrates the strong dependency of $\Delta Q$ on the quality of the
\Gaia solution.

\begin{table*}
\small
\centering
\caption{List of stars with $\Delta Q>30$ in \HTPM case A, with assumed 
radial velocities set to zero. This threshold was set for a probability of false alarm 
${\sim}10^{-5}$, assuming that $\Delta Q$ follows the $\chi^2$ distribution with 
5~degrees of freedom. The columns contain the \Hipparcos identifier, \Hipparcos magnitude, parallax, and total 
proper motion (all from the \Hipparcos Catalogue), the radial velocity from \XHIP, 
and the calculated radial motion and positional offset over $\Delta t= 23.75$~yr due 
to perspective acceleration.
\label{tab:outlierlist}}
\begin{tabular}{rrrrrrrrl} 
\toprule
HIP	&$\Delta Q$	& $Hp$	&$\varpi$& $\sqrt{\mu_{\alpha *}^2 + \mu_\delta^2}$ & $v_r$ & $\mu_r$	 
& $\Delta\theta_\text{persp}$ & Remark\\
	&		&	& [mas ] & [mas yr$^{-1}$] &  [km s$^{-1}$]	& [mas yr$^{-1}$]  &  [mas]	& \\
\cmidrule[0.2pt](lr){1-9} 	
   87937  &8\,044.46 	& 9.490	& 548.31 &10\,358.94 &$-110.51$& $-12\,782.22$& 361.87        & Barnard's star\\
   24186  &5\,053.12 	& 8.932	& 255.66 & 8\,669.40 &  245.19 &   13\,223.43 & 312.01        & Kapteyn's star\\
   57939  &   686.11 	& 6.564	& 109.99 & 7\,059.03 & $-98.35$&  $-2\,281.95$& 43.72         & Groombridge 1830\\
  104217  &   618.09 	& 6.147	& 285.88 & 5\,172.58 & $-64.07$&  $-3\,863.82$& 54.70         & 61 Cyg B\tablefootmark{a}\\
   54035  &   572.73 	& 7.506 & 392.64 & 4\,801.04 & $-84.69$&  $-7\,014.64$& 92.31         & \\
   70890  &   229.59 	&10.761	& 771.64 & 3\,852.57 & $-22.40$&  $-3\,646.21$& 38.32         & Proxima Centauri\\
   74235  &    76.86 	& 9.200	&  34.65 & 3\,681.26 &  310.12 &    2\,266.79 & 24.80         & \\
     439  &    62.69 	& 8.618	& 230.42 & 6\,100.36 &   25.38 &    1\,233.65 & 20.64         & \\
   74234  &    34.62 	& 9.568	&  35.14 & 3\,680.96 &  310.77 &    2\,303.67 & 21.63         & \\
   54211  &    30.13 	& 8.803	& 206.27 & 4\,510.10 &   68.89 &    2\,997.58 & 36.97         & \\
\bottomrule
\end{tabular}
\tablefoot{\tablefoottext{a}{61 Cyg A (HIP104214) was not included in the simulations since it is brighter than the nominal \Gaia bright star limit.}}
\end{table*}


\section{Discussion\label{sec:discussion}}

\subsection{Longevity of the \HTPM solution: detection of binary and exoplanetary candidates}

As \Gaia collects further data the accuracy of the proper motions determined from \Gaia data alone will
eventually supersede that of \HTPM. Assuming nominal mission performance and 
that the proper motion uncertainty scales with mission length as $L^{-1.5}$, this will 
happen already after 2--3~years
of \Gaia data have been accumulated. Still, \HTPM will remain a valuable source
of information as it is based on a much longer time baseline. This is relevant
for long period companions which create astrometric signatures that cannot be seen in \Gaia data alone. We
therefore suggest that \HTPM should be repeated with future \Gaia releases. The
goodness-of-fit of the combined solution is sensitive to small deviations of
the stellar motions from the assumed (rectilinear) model. This sensitivity
will dramatically increase with more \Gaia data, namely when the \Gaia-only proper
motions become as good as the combined \HTPM proper motions.

The potential for detecting faint (stellar or planetary) companions to nearby stars 
can be illustrated by a numerical example. 
Consider a $1~M_\odot$ star at 10~pc distance ($\varpi=100$~mas) from the Sun, with an invisible
companion of mass $m$ orbiting at a period of $P\simeq 25$~years (semi-major
axis $a\simeq 8.5$~au). The astrometric signature of the companion 
(i.e., the angular size of the star's orbit around their common centre of
mass; \citeads[][]{2014exha.book.....P})
is $a_\ast \simeq a\varpi (m / M_\odot) \simeq 850 (m / M_\odot)$~mas if the orbit is seen face-on, 
and the instantaneous proper motion of the star relative to the centre of mass is 
$2 \pi a_\ast / P\simeq 200 (m / M_\odot)$~mas~yr$^{-1}$.
If \Hipparcos effectively measures this instantaneous proper motion which is
extrapolated over $\Delta t = 25$~years, the extrapolated position from
\Hipparcos (with its uncertainty of about 22~mas, see Table~\ref{tab:results}) and the position observed
by \Gaia (with an uncertainty much smaller than from \Hipparcos) could differ by up
to $\simeq 5000 (m / M_\odot)$~mas. Assuming that detection is possible if the position difference 
is at least twice as large as the positional uncertainty,%
\footnote{Table~\ref{tab:outlierlist} shows that $\Delta Q$ in case~A may be sensitive to
positional deviations at the \Gaia epoch as small as 21~mas.}
we find that the initial 
\HTPM results could be sensitive to companion masses down to 
$\simeq 10^{-2}~M_\odot$, that is brown dwarf or super-Jupiter companions.

If we instead let \Gaia measure the instantaneous proper motion of the system
and propagate backwards to the \Hipparcos epoch,
we can take advantage of the much better uncertainties of the \Gaia astrometry.
Two to three years of \Gaia data already give proper motion uncertainties better 
than $30~\mu$as yr$^{-1}$
for the bright stars, and hence extrapolated position uncertainties better than
\Hipparcos at its own epoch, or $\simeq 0.75$~mas (Table~\ref{tab:results}). 
Therefore the \HTPM sensitivity increases roughly
by a factor 30, allowing the detection of companion masses down to about 
$3\times 10^{-4}~M_\odot$, or Saturn-type objects at a Saturn-like distance to the host star.

This demonstrates that the results of \HTPM can be used to find candidates for
long period exoplanets around nearby stars, with a highly interesting companion mass
range opening up with subsequent releases of \Gaia data when combined with
\Hipparcos. These companions cannot be detected from \Gaia data alone even at
the full mission length, and are hard to detect through classical methods due
to their long periods, low transit probability and small radial velocity
signatures.
Since $\Delta Q$ is sensitive to deviations from uniform space motion, whether
they are seen in the \Hipparcos or in the \Gaia data, or both, this statistic
can be used to find candidate systems in all these cases. The further exploration
of the candidate systems will however require specialised analysis tools.

In a future publication we will explore in more detail how $\Delta Q$ can be used to identify binary
and exoplanetary candidates with orbital periods of decades to centuries. 
Apart from the possibility to detect sub-stellar companions for the nearest stars, 
this will contribute to the census of the binary population within a few
hundred parsecs from the sun by filling a difficult-to-observe gap between the
shorter period spectroscopic and astrometric binaries and the visually resolved
long-period systems. 

\subsection{Two versus five parameters\label{sec:2vs5}}

When evaluating the results of our simulations, case B deserves additional
attention since it is the more realistic case for the first \Gaia data release,
and the first simulation of this case published so far. The two-parameter
solution (\Gaia12 B in Table~\ref{tab:results}) leads to a large position
error of several mas. This is caused by assuming the parallax,
proper and radial motion to be zero in the solution, whereas in reality they
are not. The actual positional uncertainties in this case depend on the true distribution 
of parallaxes and proper motions for all the stars, including the auxiliary stars, 
which are not very well known. The numerical values given here are based on 
the very schematic distribution model for the auxiliary stars described in 
Sect.~\ref{sec:simTrue}, and should therefore be interpreted with caution.

This position error is also relevant for the case B \HTPM scenario, where the
solution of the auxiliary stars is two parameters only, but where one solves all
five parameters for the \Hipparcos stars while incorporating prior information
from the \Hipparcos Catalogue. The position error of the auxiliary stars causes
a poor attitude determination. This in turn leads to increased errors in the case B \HTPM
results (compare \HTPM B and A in Table~\ref{tab:results}), with a bias in the
parallax errors (see Sect.~\ref{sec:results} and Fig.~\ref{fig:varpiHist}).
For a parallax-unbiased solution it is necessary to estimate all five
parameters for all stars included in the solution. Any mixture in the estimation of five and two parameters
in the same solution will lead to a bias in the resulting parallaxes. This is not
only true for the \HTPM scenario described in this paper but also in all
\Gaia-only data releases. 
Referring to the terminology used in Sect.~6.2 of \citetads{2012A&A...538A..78L}, 
any star for which not all five astrometric parameters can be solved must be
treated as a ``secondary source'', meaning that it does not contribute to the
attitude determination and instrument calibration. This is necessary in order 
to avoid biases for the stars where all five parameters are estimated.

\subsection{Frame rotation of the combined solution\label{sec:frame}}

For the final \AGIS solution of \Gaia the reference frame will be established
by means of quasars, both by linking to the optical counterparts of radio (VLBI)
sources defining the orientation of the International Celestial Reference Frame,
and by using the zero proper motion
of quasars to determine a non-rotating frame.\footnote{
The apparent proper motion of quasars due to the Galactocentric acceleration is expected to have an amplitude of $\sim$4~$\mu$as~yr$^{-1}$ and is taken into account when determining the spin of the reference frame.}
This can also be done for earlier \Gaia data releases, at least for the orientation
part, while the shorter time span will limit the determination of the spin.
It is desirable to rotate the \HTPM results into the same
reference frame as used for the first \Gaia data release. This must be done in two steps.
First, a provisional \HTPM must be computed in the \Hipparcos frame (as it will be when
the \Hipparcos data are used as prior, see Sect.~\ref{sec:joint}), without imposing any
other constraints on the frame. This solution will contain (many)
non-\Hipparcos stars with only \Gaia observations which include a multitude of
quasars. Their positions and proper motions are used in a second step
to correct the provisional \HTPM (and other data in the same solution) for the
estimated orientation and spin. Since the \HTPM
solution is integrated in \AGIS, the estimation and correction of the frame 
can be accomplished using the procedures and tools developed for \AGIS
(\citeads{2012A&A...538A..78L}, Sect.~6.1).


\subsection{Other applications of the joint solution method}\label{discussion:nj}
The joint solution is applicable also to other combinations of astrometric 
data. Here we give two examples.

\NJ (\citeads{2009TrSpT...7.Tm19H}; \citeads{2013IAUS..289..429Y}) is an ultra-small Japanese satellite, a
technology demonstrator for the JASMINE series of near-infrared astrometry
missions, scheduled for launch in 2015. It targets bright stars between
magnitude $1$ and $10$, although the exact limits are not yet determined. Based
on current performance estimates the uncertainties in stellar parameters will
be similar to or slightly worse than the uncertainties of the \Hipparcos data. 
However the data will still be very
valuable since astrometric catalogues are best at their respective epochs and
\NJ may be the only astrometric mission at its epoch observing the brightest
stars in the sky. The \NJ data can be analysed together with \Hipparcos data 
analogously to the \HTPM project to improve the proper motions of bright
stars that may not be observed by \Gaia \citepads{2013IAUS..289..414M}.

The \TychoTwo Catalogue \citep{2000A&A...355L..27H} gives positions for 2.5
million stars, derived from starmapper observations of \Hipparcos. The
positions at the reference epoch J1991.25 have a median internal standard error
of 7~mas for stars brighter than $V_T =
9$~mag and 60~mas for the whole catalogue. Combining the \TychoTwo positions with
\Gaia data using the joint solution scheme would allow us to derive proper
motions for these stars with median uncertainties of 0.3 and 2.5~mas~yr$^{-1}$,
respectively. This is true even in the conservative scenario (\G{12}-B), since the major
uncertainty comes from the \TychoTwo positions. In this combination the 
proper motions given in \TychoTwo should not be used, as they may contain
systematic errors of a
similar magnitude due to the incorporated old photographic material. The
derived \Tycho-\Gaia Proper Motions ({\sc tgpm}) catalogue however could be used to
correct the photographic positions in order to take advantage of the much
longer temporal baseline.

\section{Conclusions}\label{conclusions}

We have developed the joint solution method for incorporating priors in
the core astrometric solution of \Gaia.  
The method can be used in the processing of early \Gaia
data to improve the proper motions of the \Hipparcos stars, the so-called
Hundred Thousand Proper Motions project. 

Combining astrometric data from very different epochs requires careful
treatment of the non-linear effects of the mapping from spherical to
rectilinear coordinates and for high velocity stars due to perspective
acceleration. Therefore we have introduced a ``scaled model of kinematics'' (\SMOK)
which allows to handle these effects in a simple and rigorous manner.

Using simulations we have verified that \HTPM, using the joint solution method, gives
the expected large improvements in proper motion uncertainties for over 100\,000
stars in the \Hipparcos Catalogue. The predicted proper motion uncertainties
range from 14 to 134~$\mu$as~yr$^{-1}$ depending on the amount of \Gaia data used
and the stellar magnitude, about a factor 30 improvement compared with the
\Hipparcos uncertainties.

We have shown that \HTPM also delivers improved parallaxes, which however may be
strongly biased unless a full five-parameter solution can be obtained from 
\Gaia-only data also for all non-\Hipparcos stars. Whether these parallaxes
should be published as part of an \HTPM release should be decided based on the
amount and quality of \Gaia data available at the time.

The joint solution is applicable also to a combination of \TychoTwo positions
with early \Gaia data to derive improved proper motions for the 2.5 million
stars. We suggest that this possibility of a \Tycho-\Gaia Proper Motions ({\sc
tgpm}) catalogue should be considered in the \Gaia data release plan.

The proposed method to calculate \HTPM provides a goodness-of-fit measurement $\Delta
Q$ which is sensitive to deviations from the uniform linear space motion.
However, accurate radial velocities are required for nearby fast moving stars
in order to avoid mistaking outliers in $\Delta Q$ for companion signatures. We
recommend to publish $\Delta Q$ as well as the radial velocities used for
the \HTPM data reduction. This will allow further investigations of outliers
which might indicate binary or exoplanetary candidates, and will permit a correction of
the \HTPM results if better radial velocities become available.

The full power of \HTPM will not be reached with the first \Gaia data, but only in
subsequent releases benefiting from the increased sensitivity of $\Delta Q$ with
improved \Gaia results. Because of the long temporal baseline and the
combination of current with historic astrometry, \HTPM will remain relevant
throughout the final \Gaia release for the detection and measurement of binary and exoplanetary
candidates.

\begin{acknowledgements}
We thank F.~van Leeuwen for clarification on certain data items in the
\Hipparcos Catalogue and for providing valuable feedback as the referee. 
We also thank C.~Fabricius for many useful comments.
This work was partly carried out under ESA Contract No.~4000105564/12/NL/GE. 
Support from the Swedish National Space Board and the Royal Physiographic Society in Lund 
is gratefully acknowledged.
\end{acknowledgements}

\bibliographystyle{aa} 
\bibliography{HTPM,agis} 

\begin{thebibliography}{29}
\expandafter\ifx\csname natexlab\endcsname\relax\def\natexlab#1{#1}\fi

\bibitem[{Abramowitz \& Stegun(2012)}]{abramowitz2012handbook}
Abramowitz, M. \& Stegun, I. 2012, Handbook of Mathematical Functions (New
  York: Dover Publications)

\bibitem[{{Anderson} \& {Francis}(2012)}]{2012AstL...38..331A}
{Anderson}, E. \& {Francis}, C. 2012, Astronomy Letters, 38, 331

\bibitem[{{Bombrun} {et~al.}(2010){Bombrun}, {Lindegren}, {Holl}, \&
  {Jordan}}]{2010A&A...516A..77B}
{Bombrun}, A., {Lindegren}, L., {Holl}, B., \& {Jordan}, S. 2010, \aap, 516,
  A77

\bibitem[{Brinker \& Minnick(1995)}]{book:brinker+1995}
Brinker, R.~C. \& Minnick, R. 1995, {The Surveying Handbook, 2nd ed.}
  (Dordrecht: Kluwer)

\bibitem[{{de Bruijne} {et~al.}(2010){de Bruijne}, {Siddiqui}, {Lammers},
  {Hoar}, {O'Mullane}, \& {Prusti}}]{2010IAUS..261..331D}
{de Bruijne}, J., {Siddiqui}, H., {Lammers}, U., {et~al.} 2010, in IAU
  Symposium, Vol. 261, Relativity in Fundamental Astronomy: Dynamics, Reference
  Frames, and Data Analysis, ed. {S.~A.~Klioner, P.~K.~Seidelmann, \&
  M.~H.~Soffel}, 331

\bibitem[{{de Bruijne} \& {Eilers}(2012)}]{2012A&A...546A..61D}
{de Bruijne}, J.~H.~J. \& {Eilers}, A.-C. 2012, \aap, 546, A61

\bibitem[{{Dehnen} \& {Binney}(1998)}]{1998MNRAS.298..387D}
{Dehnen}, W. \& {Binney}, J.~J. 1998, \mnras, 298, 387

\bibitem[{{Dravins} {et~al.}(1999){Dravins}, {Lindegren}, \&
  {Madsen}}]{1999A&A...348.1040D}
{Dravins}, D., {Lindegren}, L., \& {Madsen}, S. 1999, \aap, 348, 1040

\bibitem[{{Eichhorn} \& {Rust}(1970)}]{eichhorn+rust1970}
{Eichhorn}, H. \& {Rust}, A. 1970, Astronomische Nachrichten, 292, 37

\bibitem[{ESA(1997)}]{hip:catalogue}
ESA. 1997, {T}he {H}ipparcos and {T}ycho {C}atalogues (Noordwijk: ESA), {E}SA
  SP-1200

\bibitem[{{Feissel} \& {Mignard}(1998)}]{1998A&A...331L..33F}
{Feissel}, M. \& {Mignard}, F. 1998, \aap, 331, L33

\bibitem[{{Hatsutori} {et~al.}(2009){Hatsutori}, {Suganuma}, {Kobayashi},
  {Gouda}, {Yano}, {Yamada}, \& {Yamauchi}}]{2009TrSpT...7.Tm19H}
{Hatsutori}, Y., {Suganuma}, M., {Kobayashi}, Y., {et~al.} 2009, Transactions
  of Space Technology Japan, 7, 19

\bibitem[{{H{\o}g} {et~al.}(2000){H{\o}g}, {Fabricius}, {Makarov}, {Urban},
  {Corbin}, {Wycoff}, {Bastian}, {Schwekendiek}, \&
  {Wicenec}}]{2000A&A...355L..27H}
{H{\o}g}, E., {Fabricius}, C., {Makarov}, V.~V., {et~al.} 2000, \aap, 355, L27

\bibitem[{{Holl} {et~al.}(2012){Holl}, {Lindegren}, \&
  {Hobbs}}]{2012A&A...543A..15H}
{Holl}, B., {Lindegren}, L., \& {Hobbs}, D. 2012, \aap, 543, A15

\bibitem[{{Lindegren} {et~al.}(2012){Lindegren}, {Lammers}, {Hobbs},
  {O'Mullane}, {Bastian}, \& {Hern{\'a}ndez}}]{2012A&A...538A..78L}
{Lindegren}, L., {Lammers}, U., {Hobbs}, D., {et~al.} 2012, \aap, 538, A78

\bibitem[{{Michalik} {et~al.}(2012){Michalik}, {Lindegren}, {Hobbs}, {Lammers},
  \& {Yamada}}]{2012ASPC..461..549M}
{Michalik}, D., {Lindegren}, L., {Hobbs}, D., {Lammers}, U., \& {Yamada}, Y.
  2012, in ASP Conference Series, Vol. 461, Astronomical Data Analysis Software
  and Systems XXI, ed. P.~{Ballester}, D.~{Egret}, \& N.~P.~F. {Lorente}, 549

\bibitem[{{Michalik} {et~al.}(2013){Michalik}, {Lindegren}, {Hobbs}, {Lammers},
  \& {Yamada}}]{2013IAUS..289..414M}
{Michalik}, D., {Lindegren}, L., {Hobbs}, D., {Lammers}, U., \& {Yamada}, Y.
  2013, in IAU Symposium, Vol. 289, Advancing the Physics of Cosmic Distances,
  ed. R.~{de Grijs}, 414

\bibitem[{Mignard(2009)}]{LL:FM-040}
Mignard, F. 2009, {T}he {H}undred {T}housand {P}roper {M}otions {P}roject,
  {Gaia} Data Processing and Analysis Consortium (DPAC) technical note
  GAIA-C3-TN-OCA-FM-040

\bibitem[{{Murray}(1983)}]{murray1983}
{Murray}, C.~A. 1983, {Vectorial astrometry} (Bristol: Adam Hilger)

\bibitem[{{Perryman}(2014)}]{2014exha.book.....P}
{Perryman}, M. 2014, {The Exoplanet Handbook} (Cambridge, UK: Cambridge
  University Press)

\bibitem[{{Risquez} {et~al.}(2013){Risquez}, {van Leeuwen}, \&
  {Brown}}]{2013A&A...551A..19R}
{Risquez}, D., {van Leeuwen}, F., \& {Brown}, A.~G.~A. 2013, \aap, 551, A19

\bibitem[{{Schlesinger}(1917)}]{1917AJ.....30..137S}
{Schlesinger}, F. 1917, \aj, 30, 137

\bibitem[{{Taff}(1981)}]{taff1981}
{Taff}, L.~G. 1981, {Computational spherical astronomy} (New York:
  Wiley-Interscience)

\bibitem[{{van Altena}(2013)}]{vanAltena2013}
{van Altena}, W.~F. 2013, {Astrometry for Astrophysics} (Cambridge, UK:
  Cambridge University Press)

\bibitem[{{van de Kamp}(1977)}]{1977VA.....21..289V}
{van de Kamp}, P. 1977, Vistas in Astronomy, 21, 289

\bibitem[{van Leeuwen(2007a)}]{book:newhip}
van Leeuwen, F. 2007a, {H}ipparcos, the {N}ew {R}eduction of the {R}aw {D}ata,
  Astrophysics and Space Science Library, Vol. 350 (Springer)

\bibitem[{{van Leeuwen}(2007b)}]{fvl2007}
{van Leeuwen}, F. 2007b, \aap, 474, 653

\bibitem[{{Wilson} \& {Hilferty}(1931)}]{1931PNAS...17..684W}
{Wilson}, E.~B. \& {Hilferty}, M.~M. 1931, Proceedings of the National Academy
  of Science, 17, 684

\bibitem[{{Yamada} {et~al.}(2013){Yamada}, {Fujita}, {Gouda}, {Kobayashi},
  {Hara}, {Nishi}, {Yoshioka}, \& {Hozumi}}]{2013IAUS..289..429Y}
{Yamada}, Y., {Fujita}, S., {Gouda}, N., {et~al.} 2013, in IAU Symposium, Vol.
  289, Advancing the Physics of Cosmic Distances, ed. R.~{de Grijs}, 429

\end{thebibliography}

\appendix

\section{Scaled Modelling of Kinematics (\SMOK)}\label{sec:smok}

A formalism called Scaled Modelling of Kinematics (\SMOK) is introduced in 
this paper to facilitate a rigorous manipulation of small (differential) quantities
in the celestial coordinates. It is reminiscent of the ``standard'' or ``tangential'' 
coordinates in classical small-field astrometry 
\citep[e.g.,][]{murray1983,vanAltena2013}, using a gnomonic projection
onto a tangent plane of the (unit) celestial sphere, but extends to three dimensions
by adding the radial coordinate perpendicular to the tangent plane. This simplifies 
the modelling of perspective effects.

Figure~\ref{fig:smok2}  illustrates the concept. In the vicinity of the 
star let $\vec{c}$ be a comparison point fixed with respect to the Solar System 
Barycentre (SSB). As shown in the diagrams:
\begin{enumerate}
\item
The barycentric motion of the star is scaled by the inverse distance to $\vec{c}$, 
effectively placing the star on or very close to the unit sphere. 
\item
Rectangular coordinates are expressed in the barycentric $[\vec{p}_\text{c}~\vec{q}_\text{c}~\vec{r}_\text{c}]$ system
with $\vec{r}_\text{c}$ pointing towards $\vec{c}$, and $\vec{p}_\text{c}$, $\vec{q}_\text{c}$ 
in the directions of increasing right ascension and declination.
\end{enumerate}
The first point eliminates the main uncertainty in the kinematic modelling of the star due to
its poorly known distance. The second point allows us to express the scaled kinematic
model in \SMOK coordinates $a$, $d$, $r$ that are locally aligned with $\alpha$, $\delta$, and 
the barycentric vector.

\begin{figure}
\centering
\setlength{\fboxsep}{5pt}\setlength{\fboxrule}{0.5pt}
\fbox{
\includegraphics[width=0.8\columnwidth,clip=true]{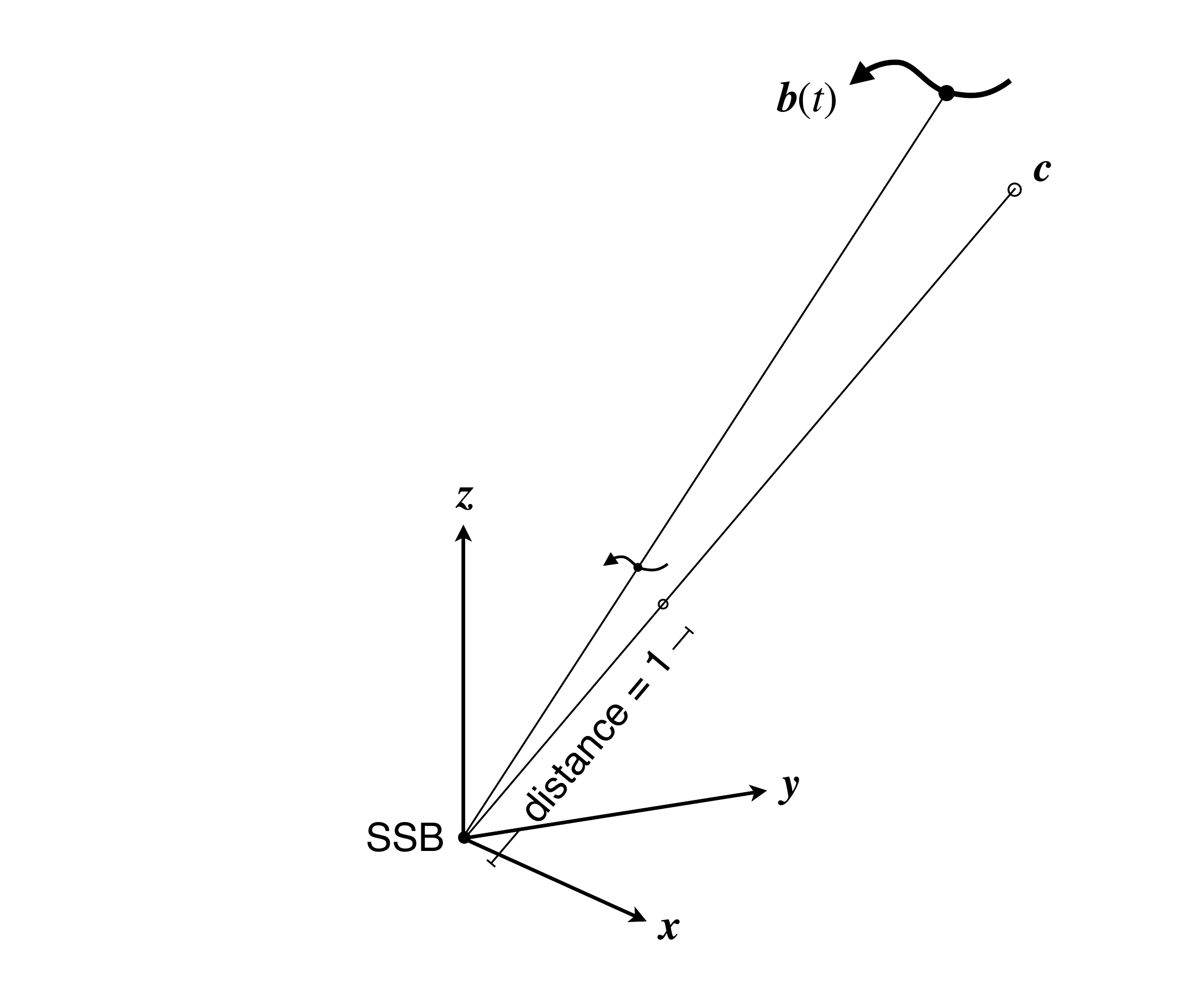}
}
\centerline{\vspace{2mm}}
\fbox{
\includegraphics[width=0.8\columnwidth,clip=true]{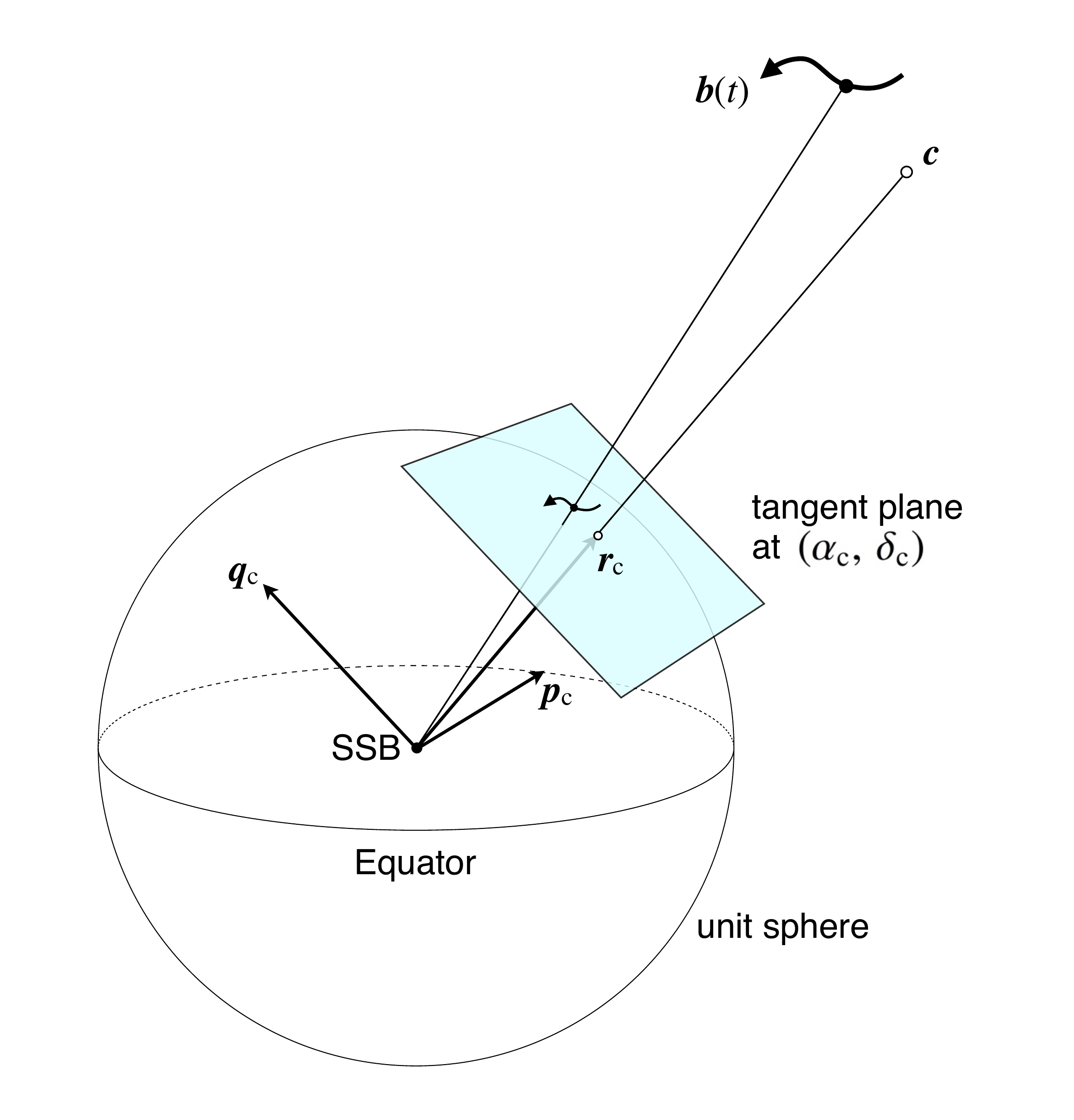}
}
\caption{Two steps in the definition of \SMOK coordinates.
In the top diagram the motion of an object in the vicinity 
of the fixed point $\vec{c}$ is modelled by the function $\vec{b}(t)$ 
expressed in the barycentric $[\vec{x}~\vec{y}~\vec{z}]$ system. 
A scaled version of the model is constructed such that the scaled $\vec{c}$ 
is at unit distance from the Solar System Barycentre (SSB).
In the bottom diagram new coordinate axes $[\vec{p}_\text{c}~\vec{q}_\text{c}~\vec{r}_\text{c}]$ are 
chosen in the directions of increasing right ascension, declination, and distance,
respectively, at the comparison point $(\alpha_\text{c},\,\delta_\text{c})$ being 
the projection of $\vec{c}$ on the unit sphere. 
\label{fig:smok2}}
\end{figure}

Up to the scale factor $|\vec{c}|^{-1}$ discussed below, the \SMOK coordinate system 
is completely defined by the adopted comparison point $(\alpha_\text{c},\,\delta_\text{c})$
using the orthogonal unit vectors
\begin{equation}\label{eq1}
\vec{p}_\text{c} =
\begin{bmatrix} -\sin\alpha_\text{c} \\ \phantom{-}\cos\alpha_\text{c} \\ 0 \end{bmatrix} ,
\quad
\vec{q}_\text{c} =
\begin{bmatrix} -\sin\delta_\text{c}\cos\alpha_\text{c} \\ -\sin\delta_\text{c}\sin\alpha_\text{c} \\
\cos\delta_\text{c} \end{bmatrix} ,
\quad
\vec{r}_\text{c} =
\begin{bmatrix} \cos\delta_\text{c}\cos\alpha_\text{c} \\ \cos\delta_\text{c}\sin\alpha_\text{c} \\
\sin\delta_\text{c} \end{bmatrix} .
\end{equation}
$[\vec{p}_\text{c}~\vec{q}_\text{c}~\vec{r}_\text{c}]$ is the ``normal triad'' at the comparison point with respect to 
the celestial coordinate system \citep{murray1983}.%
\footnote{$\vec{p}_\text{c}$ and $\vec{q}_\text{c}$ point to the local ``East'' and ``North'', respectively, provided 
that $|\delta_\text{c}|<90^\circ$. However, the coordinate triad in Eq.~(\ref{eq1}) is well-defined 
even \emph{exactly} at the poles, where $\alpha_\text{c}$ remains significant for defining
$\vec{p}_\text{c}$ and $\vec{q}_\text{c}$.}
We are free to choose $(\alpha_\text{c},\,\delta_\text{c})$ as it will best serve our purpose,
but once chosen (for a particular application) it is fixed: it has no proper motion, no parallax, and 
no associated uncertainty.
Typically $(\alpha_\text{c},\,\delta_\text{c})$ is chosen very close to the mean position
of the star. 

The motion of the star in the Barycentric Celestial Reference System (BCRS) is represented
by the function $\vec{b}(t)$, where $\vec{b}$ is the vector from SSB to the star as it would be
observed from the SSB at time $t$.
The scaled kinematic model $\vec{s}(t)=\vec{b}(t)|\vec{c}|^{-1}$ 
is given in \SMOK coordinates as
\begin{equation}\label{eq2}
a(t) = \vec{p}_\text{c}'\vec{s}(t)\, , \quad
d(t) = \vec{q}_\text{c}'\vec{s}(t)\, , \quad
r(t) = \vec{r}_\text{c}'\vec{s}(t)\, ,
\end{equation} 
and can in turn be reconstructed from the \SMOK coordinates as
\begin{equation}\label{eq3}
\vec{s}(t) = \vec{p}_\text{c} a(t) + \vec{q}_\text{c} d(t) + \vec{r}_\text{c} r(t) \, .
\end{equation}
$a$, $d$, $r$ are dimensionless and the first two are typically small quantities 
($\lesssim 10^{-4}$), while $r$ is very close to unity. 

The whole point of the scaled kinematic modelling is that $\vec{s}(t)$ can be described
very accurately by astrometric observations, even though $\vec{b}(t)$ may be poorly 
known due to a large uncertainty in distance. 
This is possible simply by choosing the scaling such that $|\vec{s}(t)|=1$ at some suitable time. 
This works even if the distance is completely 
unknown, or if it is effectively infinite (as for a quasar).

The scale factor is $|\vec{c}|^{-1}=\varpi_\text{c}/A$, where $\varpi_\text{c}$ is the 
parallax of $\vec{c}$ and $A$ the astronomical unit. The measured parallax 
can be regarded as an estimate of $\varpi_\text{c}$.

In the following we describe some typical applications of \SMOK coordinates.

\subsection{Uniform space motion}

The simplest kinematic model is to assume that the star moves uniformly with respect
to the SSB, that is
\begin{equation}\label{eq4}
\vec{b}(t)=\vec{b}_\text{ep} + (t-t_\text{ep})\vec{v} \, ,
\end{equation}  
where $\vec{b}_\text{ep}$ is the barycentric position at the reference epoch 
$t_\text{ep}$, and $\vec{v}$ is the (constant) space velocity. The scaled kinematic
model expressed in the BCRS is
\begin{equation}\label{eq4s}
\vec{s}(t)=\vec{s}_\text{ep} + (t-t_\text{ep})\vec{\dot{s}} \, ,
\end{equation} 
where
\begin{equation}\label{eq4sa}
\vec{s}_\text{ep} = \vec{p}_\text{c} a(t_\text{ep}) + \vec{q}_\text{c} d(t_\text{ep}) + \vec{r}_\text{c} r(t_\text{ep}) \, 
\end{equation}
and
\begin{equation}\label{eq4sb}
\vec{\dot{s}} = \vec{p}_\text{c} \dot{a} + \vec{q}_\text{c} \dot{d} + \vec{r}_\text{c} \dot{r} \, 
\end{equation}
are constant vectors. The uniform motion can also be written in \SMOK coordinates as
\begin{equation}\label{eq5}
\left.
\begin{aligned}
a(t) &= a(t_\text{ep}) + (t-t_\text{ep})\dot{a}\, , \\
d(t) &= d(t_\text{ep}) + (t-t_\text{ep})\dot{d}\, , \\
r(t) &= r(t_\text{ep}) + (t-t_\text{ep})\dot{r}\, .
\end{aligned} \quad \right\} 
\end{equation}
The six constants $a(t_\text{ep})$, $d(t_\text{ep})$, $r(t_\text{ep})$, $\dot{a}$, 
$\dot{d}$, $\dot{r}$ are the kinematic parameters of the scaled model; however, 
to get the actual kinematics of the star we also need to know $\varpi_\text{c}$.  

\subsection{Relation to the usual astrometric parameters}

Choosing $(\alpha_\text{c},\,\delta_\text{c})$ to be the barycentric celestial
coordinates of the star at $t_\text{ep}$, and $\varpi_\text{c}$ equal to the parallax
at the same epoch, we find
\begin{equation}\label{eq6}
\left.
\begin{aligned}
a(t_\text{ep})&=0\, , & d(t_\text{ep})&=0\, , & r(t_\text{ep})&=1\, , \\ 
\dot{a}&=\mu_{\alpha*}\, ,  & \dot{d}&=\mu_\delta\, , & \dot{r}&=\mu_r\, ,
\end{aligned}\quad
\right\} 
\end{equation}  
where $\mu_{\alpha*}$, $\mu_\delta$ are the tangential components of the barycentric
proper motion at the reference epoch $t_\text{ep}$, and $\mu_r$
is the ``radial proper motion'' allowing to take the perspective effects into account. 
$\mu_r$ is usually calculated from the measured 
radial velocity and parallax according to Eq.~(\ref{eq:muR}).

\subsection{Differential operations}

Uniform space motion does not map into barycentric coordinates 
$\alpha(t)$, $\delta(t)$ that are linear functions of time. The non-linearity derives 
both from the curvilinear nature of spherical coordinates and from perspective
foreshortening depending on the changing distance to the object. Both effects are 
well known and have been dealt with rigorously by several authors 
\citep[e.g.,][]{eichhorn+rust1970,taff1981}. The resulting expressions are 
non-trivial and complicate the comparison of astrometric catalogues of different 
epochs. For example, approximations such as
\begin{equation}\label{eq7}
\mu_{\alpha*} = \frac{\alpha(t_2)-\alpha(t_1)}{t_2-t_1}\cos\delta\, , \quad
\mu_\delta = \frac{\delta(t_2)-\delta(t_1)}{t_2-t_1}
\end{equation}
cannot be used when the highest accuracy is required. By contrast, the linearity 
of Eq.~(\ref{eq5}) makes it possible to write
\begin{equation}\label{eq8}
\dot{a} = \frac{a(t_2)-a(t_1)}{t_2-t_1}\, , \quad
\dot{d} = \frac{d(t_2)-d(t_1)}{t_2-t_1}
\end{equation}
to full accuracy, provided that the same comparison point is used for both
epochs. (Strictly speaking, the same scale factor must also be used, so that
in general $r(t_2)-r(t_1)=(t_2-t_1)\dot{r}\ne 0$.) If the position at the 
reference epoch coincides with the comparison point used, the resulting 
$\dot{a}$, $\dot{d}$ are the looked-for proper motion components according
to Eq.~(\ref{eq6}); otherwise a change of comparison point is needed (see below). 
 
\subsection{Changing the comparison point}\label{sec:changecp}

Let $(\alpha_1,\,\delta_1)$ and $(\alpha_2,\,\delta_2)$ be different comparison
points with associated triads $[\vec{p}_1~\vec{q}_1~\vec{r}_1]$ and
$[\vec{p}_2~\vec{q}_2~\vec{r}_2]$. If $a_1(t)$, $d_1(t)$, $r_1(t)$ and
$a_2(t)$, $d_2(t)$, $r_2(t)$ describe the same scaled kinematics we have
by Eq.~(\ref{eq3})
\begin{equation}\label{eq9}
\vec{s}(t) =
\vec{p}_1 a_1(t) + \vec{q}_1 d_1(t) + \vec{r}_1 r_1(t) =
\vec{p}_2 a_2(t) + \vec{q}_2 d_2(t) + \vec{r}_2 r_2(t)\, .
\end{equation}
Thus, given $a_1(t)$, $d_1(t)$, $r_1(t)$ one can compute $\vec{s}(t)$ from the 
first equality in Eq.~(\ref{eq9}), whereupon the modified functions are recovered as 
\begin{equation}\label{eq10}
a_2(t) = \vec{p}_2'\vec{s}(t)\, , \quad
d_2(t) = \vec{q}_2'\vec{s}(t)\, , \quad
r_2(t) = \vec{r}_2'\vec{s}(t)\, .
\end{equation}
This procedure can be applied to $\vec{s}(t)$ for any particular $t$ as well as to
linear operations on $\vec{s}$ such as differences and time derivatives.

\subsection{Epoch propagation}\label{sec:propag}

An important application of the above formulae is for propagating the six astrometric 
parameters $(\alpha_1,\,\delta_1,\,\varpi_1,\,\mu_{\alpha*1},\,\mu_{\delta 1},\,\mu_{r1})$,
referring to epoch $t_1$, to a different epoch $t_2$. This can be done in the following steps:
\begin{enumerate}
\item
Use $(\alpha_1,\,\delta_1)$ as the comparison point
and compute $[\vec{p}_1~\vec{q}_1~\vec{r}_1]$ by Eq.~(\ref{eq1}). At time $t_1$ the 
\SMOK parameters relative to the first comparison point are $a_1(t_1)=d_1(t_1)=0$, 
$r_1(t_1)=1$, $\dot{a}_1=\mu_{\alpha*1}$, $\dot{d}_1=\mu_{\delta 1}$, $\dot{r}_1=\mu_{r1}$.
\item
Calculate $\vec{s}(t_1)$ and $\vec{\dot{s}}$ using Eqs.~(\ref{eq4sa})--(\ref{eq4sb}).
\item
Calculate $\vec{s}(t_2)$ by means of Eq.~(\ref{eq4s}).
Let $s_2=|\vec{s}(t_2)|$ be its length (close to unity).
\item
Calculate $\vec{r}_2=\vec{s}(t_2)/s_2$ and hence the second comparison
point $(\alpha_2,\,\delta_2)$ and triad $[\vec{p}_2~\vec{q}_2~\vec{r}_2]$.
\item
Use Eq.~(\ref{eq10}) to calculate the \SMOK parameters at $t_2$ referring to
the second comparison point. For the position one trivially gets $a_2(t_2)=d_2(t_2)=0$ 
and $r_2(t_2)=s_2$. For the proper motion parameters one finds
$\dot{a}_2=\vec{p}_2'\vec{\dot{s}}$, $\dot{d}_2=\vec{q}_2'\vec{\dot{s}}$, and 
$\dot{r}_2=\vec{r}_2'\vec{\dot{s}}$.
\item
The astrometric parameters at epoch $t_2$ are $\alpha_2$, $\delta_2$, 
$\varpi_2=\varpi_1/s_2$, $\mu_{\alpha*2}=\dot{a}_2/s_2$, $\mu_{\delta 2}=\dot{d}_2/s_2$, 
$\mu_{r 2}=\dot{r}_2/s_2$. 
\end{enumerate} 
This procedure is equivalent to the one described in Sect.~1.5.5, Vol.~1 of  
\emph{The Hipparcos and Tycho Catalogues} \citep{hip:catalogue}. 

\section{The \Hipparcos Catalogue}\label{sec:hip2}

This Appendix describes the calculation of relevant quantities from the new reduction of the \Hipparcos
Catalogue by \citet{fvl2007}. Data files were retrieved from the Strasbourg astronomical Data Center
(CDS) in November 2013 (catalogue I/311). These files differ slightly from the ones given on the DVD 
published along with the book \citep{book:newhip}, both in content and format, as some errors have 
been corrected. The data needed for every accepted catalogue entry are:
\begin{itemize}
\item the five astrometric parameters $(\alpha,~\delta,~\varpi,~\mu_{\alpha*},~\mu_\delta)$;
\item the $5\times5$ normal matrix $\vec{N}$ from the least-squares solution of the 
astrometric parameters (for a 5-parameter solution this equals the inverse of the 
covariance matrix $\vec{C}$);
\item the chi-square goodness-of-fit quantity $Q$ for the 5-parameter solution of the 
\Hipparcos data;
\item the degrees of freedom $\nu$ associated with $Q$.
\end{itemize}
The astrometric parameters at the \Hipparcos reference epoch J1991.25 are directly taken from
the fields labelled {\tt RArad}, {\tt DErad}, {\tt Plx}, {\tt pmRA}, and {\tt pmDE} in the main catalogue
file {\tt hip2.dat}. Units are [rad] for $\alpha$ and $\delta$, [mas] for $\varpi$, and [mas~yr$^{-1}$]
for $\mu_{\alpha*}$ and $\mu_\delta$. It is convenient to express also positional differences (such 
as \SMOK coordinates $a$ and $d$) and positional uncertainties in [mas]. The elements of $\vec{N}$
thus have units [mas$^{-2}$~yr$^{\,p}$], where $p=0$, 1, or 2, depending on the position of the
element in the matrix.

The calculation of $\vec{N}$, $Q$, and $\nu$ is described hereafter in some detail as the
specification of $\vec{C}$ deviates in some details from the published documentation. Clarification 
on certain issues was kindly provided by F.~van Leeuwen (private comm.).

The number of degrees of freedom is
\begin{equation}\label{hip2e01}
\nu = N_\text{tr} - n \, ,
\end{equation}
where $N_\text{tr}$ is the number of field transits used (label {\tt Ntr} in {\tt hip2.dat}) and
$n$ is the number of parameters in the solution (see below; most stars have $n=5$). The
goodness-of-fit given in field {\tt F2} is the ``gaussianized'' chi-square
\citepads{1931PNAS...17..684W}
\begin{equation}\label{hip2e02}
F_2 = \left( \frac{9\nu}{2} \right)^{1/2} \left[ \left( \frac{Q}{\nu} \right)^{1/3} + \frac{2}{9\nu} - 1 \right]
\end{equation}
computed from $Q$, the sum of the squared normalized residuals, and $\nu$. For ``good'' 
solutions $Q$ is expected to follow the chi-square distribution with $\nu$ degrees of freedom
($Q\sim\chi^2(\nu)$), in which case $F_2$ approximately follows the standard normal 
distribution, $F_2\sim N(0,1)$. Thus, $F_2>3$ means that $Q$ is ``too large'' for the given 
$\nu$ at the same level of significance as the $+3\sigma$ criterion for a Gaussian variable
(probability $\la 0.0044$).%
\footnote{This transformation was also used to generate the F2 statistic given in field H30 
of the \Hipparcos and \Tycho Catalogues \citep{hip:catalogue}.}
Given $F_2$ from field {\tt F2}, and $\nu$ from Eq.~(\ref{hip2e01}), it is therefore possible 
to reconstruct the chi-square statistic of the $n$-parameter solution as
\begin{equation}\label{hip2e03}
Q = \nu\left[ \left( \frac{2}{9\nu}\right)^{1/2}F_2 + 1 - \frac{2}{9\nu} \right]^3 \, .
\end{equation} 
We also introduce the square-root of the reduced chi-square,
\begin{equation}\label{hip2e04}
u = \sqrt{Q/\nu} \, ,
\end{equation} 
which is expected to be around 1.0 for a ``good'' solution (see further discussion below).
$u$ is sometimes referred to as the standard error of unit weight \citep{book:brinker+1995}.

The catalogue gives the covariance matrix in the form of an upper-diagonal 
``weight matrix'' $\vec{U}$ such that, formally, $\vec{C}=(\vec{U}'\vec{U})^{-1}$. 
This inverse exists for all stars where a solution is given. (For the joint solution
we actually need the normal matrix $\vec{N}=\vec{U}'\vec{U}$, see below.) 
For solutions with $n=5$ 
astrometric parameters there are $n(n+1)/2=15$ non-zero elements in $\vec{U}$. 
For some stars the solution has more than five parameters, and the main catalogue
then only gives the first 15 non-zero elements, while remaining elements are given in 
separate tables. Let $U_1$, $U_2$, $\dots$, $U_{15}$ be the 15 values taken from the fields labelled 
{\tt UW} in {\tt hip2.dat}. The matrix $\vec{U}$ is computed as
\begin{equation}\label{hip2e05}
\vec{U} = \begin{bmatrix} f_1U_1 & U_2 & U_4 & U_7 & U_{11} \\
0 & f_2U_3 & U_5 & U_8 & U_{12} \\
0 & 0 & f_3U_6 & U_9 & U_{13}\\
0 & 0 & 0 & f_4U_{10} & U_{14}\\
0 & 0 & 0 & 0 & f_5U_{15}\\ \end{bmatrix} \, .
\end{equation}
Here $f_i$, $i=1\dots n$, are scaling factors which for the CDS data must be calculated as
\begin{align}\label{hip2e06}
f_1 &= u/\sigma_{\alpha*}\, , &
f_2 &= u/\sigma_{\delta}\, , &
f_3 &= u/\sigma_{\varpi}\, ,  \nonumber \\ 
f_4 &= u/\sigma_{\mu\alpha*}\, , & 
f_5 &= u/\sigma_{\mu\delta}\, ,
\end{align}
where $u$ is given by Eq.~(\ref{hip2e04}) and $\sigma_{\cdot}$ are the standard errors 
given in fields {\tt e\_RArad} through {\tt e\_pmDE} of {\tt hip2.dat}. Equation~(\ref{hip2e06})
applies to data taken from the CDS version of the catalogue (I/311). For catalogue data on the 
DVD accompanying the book \citep{book:newhip}, scaling factors $f_i=1$ apply, although 
those data are superseded by the CDS version.

The $5\times 5$ matrix $\vec{N}=\vec{U}'\vec{U}$ computed using the first five rows and 
columns in $\vec{U}$, as given in Eq.~(\ref{hip2e05}), contains the relevant elements of the
normal matrix for any solution with $n\ge 5$. Thus, for solutions with $n>5$ there is no
need, for the catalogue combination, to retrieve the additional elements of $\vec{U}$ from
{\tt hip7p.dat}, etc. The situation is different when the covariance matrix is needed: it is 
then necessary to compute the full $n\times n$ normal matrix $\vec{N}$ before  
$\vec{C}=\vec{N}^{-1}$ can be computed.

The normal matrix $\vec{N}$ computed as described above incorporates the formal
uncertainties of the observations; as described in \citet{book:newhip} these are ultimately 
derived from the photon statistics of the raw data after careful analysis of the residuals as 
function of magnitude, etc. If the adopted models are correct we expect the $F_2$ statistic 
to be normally distributed with zero mean and unit standard deviation, and the standard 
error of unit weight, $u$, to be on the average equal to 1. In reality we find (for solutions 
with $n=5$) 
that their distributions are skewed towards larger values, especially for the bright stars
where photon noise is small and remaining calibration errors are therefore relatively more 
important. 
To account for such additional errors the published standard errors $\sigma_{\alpha*}$, 
etc., in {\tt hip2.dat} include, on a star-by-star basis, a correction factor equal to the 
unit weight error $u$ obtained in its solution. This is equivalent to scaling the formal
standard errors of the data used in the solution by the same factor.  In order to make 
the computed normal matrix, covariance matrix, and goodness-of-fit statistics 
consistent with the published standard errors it is then necessary to apply the 
corresponding corrections, viz.:
\begin{equation}\label{hip2e07}
\vec{N}_\text{corr} = \vec{N} u^{-2} \, , \quad
\vec{C}_\text{corr} = \vec{C} u^2 \, , \quad
Q_\text{corr} = \nu \, , \quad
u_\text{corr} = 1 \, .
\end{equation}
For the catalogue combination we use $\vec{N}_\text{corr}$ and $Q_\text{corr}$ 
whenever $u>1$, but $\vec{N}$ and $Q$ if $u\le 1$.  
 
\end{document}